\documentstyle[12pt]{article}
\newcommand{\be}{\begin{equation}}
\newcommand{\ee}{\end{equation}}
\newcommand{\bea}{\begin{eqnarray}}
\newcommand{\eea}{\end{eqnarray}}
\newcommand{\norsl}{\normalsize\sl}
\newcommand{\norsc}{\normalsize\sc}
\begin{document}

\begin{titlepage}

\title{THE BACKGROUND FIELD METHOD:
       Alternative Way of Deriving the Pinch Technique's Results}
\author{
\norsc  Shoji Hashimoto\thanks{Supported in part by the Monbusho
          Grant-in-Aid for Scientific Research No. 040011.}
         \,, Jiro KODAIRA\thanks{Supported in part by the Monbusho
                Grant-in-Aid for Scientific Research
           No. C-05640351.} \,and Yoshiaki YASUI
         \thanks{Supported in part by the Monbusho
          Grant-in-Aid for Scientific Research No. 050076.}\\
\norsl  Dept. of Physics, Hiroshima University\\
\norsl  Higashi-Hiroshima 724, JAPAN\\
\\
\norsc  Ken SASAKI\thanks{e-mail address: a121004@c1.ed.ynu.ac.jp}\\
\norsl  Dept. of Physics, Yokohama National University\\
\norsl  Yokohama 240, JAPAN\\}

\date{}
\maketitle
 
\begin{abstract}
{\normalsize We show that the background field method (BFM) is a simple 
way of deriving the same gauge-invariant results which are obtained 
by the pinch technique (PT). For illustration we construct
gauge-invariant self-energy and three-point vertices for gluons
at one-loop level by BFM and demonstrate that 
we get the same results which were derived via 
PT. We also calculate the four-gluon vertex in BFM and show 
that this vertex obeys the same Ward identity that was found with PT.}
\end{abstract}
 
\begin{picture}(5,2)(-290,-600)
\put(2.3,-110){HUPD-9408}
\put(2.3,-125){YNU-HEPTh-94-104}
\put(2.3,-140){May 1994}
\end{picture}

\thispagestyle{empty}
\end{titlepage}
\setcounter{page}{1}
\baselineskip 20pt
\section{Introduction}
\smallskip

Formulation of a gauge theory begins with a gauge invariant 
Lagrangian. However, except for lattice gauge theory, when we 
quantize the theory in the continuum we are under compulsion to fix a gauge. 
Consequently, the corresponding Green's functions, in general, will not be 
gauge invariant. These Green's functions in the standard formulation do not 
directly reflect the underlying gauge invariance of the theory but rather 
obey complicated Ward identities. If there is a method in which we can 
construct systematically gauge-invariant Green's functions, then it will 
make the computations much simpler and may have many applications.

Along this line exist two approaches: One is the pinch technique and 
the other, the background field method. The pinch technique (PT) was 
proposed some time ago by 
Cornwall~\cite{rCa}~\cite{rCb} for a well-defined algorithm to 
form new gauge-independent proper vertices and new propagators 
with gauge-invariant self-energies. Using this technique 
Cornwall and Papavassiliou obtained the one-loop 
gauge-invariant self-energy and vertex parts in 
QCD~\cite{rCP}~\cite{rPa}. Later it was shown~\cite{rPb} 
that PT works also 
in spontaneously broken gauge theories, and since then it has been 
applied to the standard model to obtain a gauge-invariant electromagnetic 
form factor of the neutrino~\cite{rPb}, one-loop gauge-invariant 
$WW$ and $ZZ$ self-energies~\cite{rDS},  
and $\gamma WW$ and $ZWW$ vertices~\cite{rPP}.

On the other hand, the background field method (BFM) was first
introduced by DeWitt~\cite{Dea} as a technique for quantizing gauge 
field theories while retaining explicit gauge invariance. In 
its original formulation, DeWitt worked only for 
one-loop calculations. The multi-loop extension of the method was given by 
't Hooft~\cite{tH}, DeWitt~\cite{Deb}, Boulware~\cite{Bo}, and 
Abbott~\cite{Abb}. Using these extensions of the background field method, 
explicit two-loop calculations of the $\beta$ function for pure Yang-Mills 
theory was made first in the Feynman gauge~\cite{Abb}-~\cite{IO}, and later 
in the general gauge~\cite{CM}. 

Both PT and BFM have the same interesting feature. The Green's functions 
(gluon self-energies and proper gluon-vertices, etc) 
constructed by the two methods retain the explicit gauge invariance, thus 
obey the naive Ward identities. As a result, for example, a computation of 
the QCD $\beta$-function coefficient is much simplified. 
Only thing we need to do is to construct the 
gauge-invariant gluon self-energy in either method and to examine 
its ultraviolet-divergent part. 
Either method gives the same correct answer~\cite{rCP}~\cite{Abb}. 
Thus it may be plausible to anticipate that 
PT and BFM are equivalent and that they produce exactly the same results.

In this paper we show that BFM is an alternative and 
simple way of deriving the same 
gauge-invariant results which are obtained by PT. Although the final results 
obtained by both methods are gauge invariant, we have found, in particular,   
that the BFM in the Feynman gauge corresponds to the intrinsic PT.
In fact we explicitly demonstrate, for the cases of the 
gauge-invariant gluon self-energy and three-point vertex, both 
methods with the Feynman gauge produce the same results 
which are equal term by term. We also give the gauge-invariant 
four-gluon vertex calculated in BFM and show explicitly
that this vertex satisfies the 
same simple Ward identity that was found with PT.

The paper is organized as follows. In Sec.2 we review the intrinsic PT 
and explain how the gauge-invariant proper self-energy and 
three-point vertex for gluon were derived in PT. In Sec.3 we write down the 
Feynman rule for QCD in BFM and compute the gauge-invariant gluon 
self-energy at one-loop level in BFM with the Feynman gauge. 
The result is shown to be the 
same, term by term, as the one obtained by the intrinsic PT. 
The BFM is applied, in Sec.4, to the calculation of 
the three-gluon vertex.  
The result is shown to coincide, again term by term, with 
the one derived by the intrinsic PT. In Sec.5 we compute the gauge-invariant 
four-gluon vertex at one-loop level in BFM. We present each contribution to 
the vertex from the individual Feynman diagram.  
Then we show that the acquired four-gluon vertex satisfies the same naive Ward 
identity that was found with PT and is related to the gauge-invariant 
three-gluon vertex obtained previously by PT and BFM.

\bigskip 
\section{The intrinsic pinch technique}
\smallskip

There are three equivalent versions of the pinch technique: the $S$ matrix 
PT~\cite{rCa} -~\cite{rCP}, the intrinsic PT~\cite{rCP} and 
the Degrassi-Sirlin alternative formulation of the PT~\cite{rDS}. 
To prepare for the later discussions and to establish the notations, 
we briefly review, in this section, the intrinsic PT and explain the way 
how the gauge-invariant proper self-energy and three-point vertex for gluons 
at one-loop level were obtained in Ref.~\cite{rCP}. 

In the $S$-matrix pinch technique we obtain the gauge-invariant effective  
gluon propagator by adding the pinch graphs in Fig. 1(b) and 1(c) 
to the ordinary propagator graphs [Fig. 1(a)]. 
The gauge dependence of the ordinary 
graphs is canceled by the contributions from the pinch graphs. 
Since the pinch graphs are always missing one or more 
propagators corresponding to the external legs, the gauge-dependent parts
of the ordinary graphs must also 
be missing one or more external propagator legs. 
So if we extract systematically from the proper graphs the parts  
which are missing external propagator legs and simply throw them away,  
we obtain the gauge-invariant results. 
This is the intrinsic PT introduced by Cornwall and Papavassiliou~\cite{rCP}. 

We will now derive the gauge-invariant proper self-energy for gluons 
of gauge group $SU(N)$ using the intrinsic PT.
Since we know that PT successfully gives gauge-invariant quantities, 
we use the Feynman gauge. 
Then the ordinary proper self-energy 
whose corresponding graphs are shown in Fig. 2 is given by
\bea
     \Pi^{0}_{\mu \nu} &=& \frac{iNg^{2}}{2} \int 
               \frac{d^D k}{(2\pi)^D}\frac{1}{k^2 (k+q)^2} \nonumber \\
     & & \qquad \times [\Gamma_{\alpha \mu \lambda}(k,q) 
                         \Gamma_{\lambda \nu \alpha}(k+q,-q) 
             - k_{\mu}(k+q)_{\nu}-k_{\nu}(k+q)_{\mu}],
\label{SelfE}
\eea
where we have symmetrized the ghost loop in Fig. 2(b) and omitted a trivial
group-theoretic factor $\delta^{ab}$. We assume the dimensional
regularization in $D=4 - 2\varepsilon $ dimensions. The three-gluon vertex 
$\Gamma_{\alpha \mu \lambda}(k,q)$ has the following expression~\cite{defGamma}:
\bea
  \Gamma_{\alpha \mu \lambda}(k,q) &\equiv& 
              \Gamma_{\alpha \mu \lambda}(k,q,-k-q)  \nonumber \\
    &=&(k-q)_{\lambda}g_{\alpha \mu}+(k+2q)_{\alpha}g_{\mu \lambda}
              -(2k+q)_{\mu}g_{\lambda \alpha}.
\label{Gam}
\eea
Here and in the following we make it a rule that whenever the external 
momentum appears in the three-gluon vertex, we put it in the middle of the 
expression, that is, like $q_{\mu}$ in Eq. (\ref{Gam}).
Now we decompose the vertices into two pieces: a piece $\Gamma^{F}$ which 
has the terms with the external momentum $q$ and a piece 
$\Gamma^{P}$ ($P$ for pinch) carries the internal momenta only.
\bea
  \Gamma_{\alpha \mu \lambda}(k,q) &=& 
       \Gamma^{F}_{\alpha \mu \lambda}+\Gamma^{P}_{\alpha \mu \lambda}, 
  \nonumber \\
  \Gamma^{F}_{\alpha \mu \lambda}(k,q) &=&
        -(2k+q)_{\mu}g_{\lambda \alpha}
        +2q_{\alpha}g_{\mu \lambda}-2q_{\lambda}g_{\alpha \mu}, \label{FGGG} \\ 
  \Gamma^{P}_{\alpha \mu \lambda}(k,q) &=& 
         k_{\alpha}g_{\mu \lambda}+(k+q)_{\lambda}g_{\alpha \mu}. \nonumber 
\eea
The full vertex $\Gamma_{\alpha \mu \lambda}(k,q)$ satisfies the 
following Ward identities:
\bea
    k^{\alpha}\Gamma_{\alpha \mu \lambda}(k,q) &=& 
         P_{\mu \lambda}(q) d^{-1}(q) - 
          P_{\mu \lambda}(k+q) d^{-1}(k+q)   \nonumber \\ 
   (k+q)^{\lambda}\Gamma_{\alpha \mu \lambda}(k,q) &=& 
         P_{\alpha \mu}(q) d^{-1}(q) - 
             P_{\alpha \mu}(k) d^{-1}(k)  
\label{Ward}
\eea
where we have defined 
\be
    P_{\mu \nu}(q) = -g_{\mu \nu} + q_{\mu}q_{\nu}q^{-2}, \qquad \quad 
          d^{-1}(q) = q^2 .
\ee

The rules of the intrinsic PT are to let the pinch vertex $\Gamma^{P}$ 
act on the full vertex and to throw out the $d^{-1}(q)$ terms thereby 
generated. We rewrite the product of two full vertices 
$\Gamma_{\alpha \mu \lambda} \Gamma_{\lambda \nu \alpha}$ as
\be
 \Gamma_{\alpha \mu \lambda} \Gamma_{\lambda \nu \alpha}=
    \Gamma^{F}_{\alpha \mu \lambda} \Gamma^{F}_{\lambda \nu \alpha}+
    \Gamma^{P}_{\alpha \mu \lambda} \Gamma_{\lambda \nu \alpha}+
    \Gamma_{\alpha \mu \lambda} \Gamma^{P}_{\lambda \nu \alpha}-
    \Gamma^{P}_{\alpha \mu \lambda} \Gamma^{P}_{\lambda \nu \alpha}
\label{Gamb}
\ee
Using the Ward identities in Eq. (\ref{Ward}) we find that the sum of 
the second and third terms of Eq. (\ref{Gamb}) 
turns out to be
\bea
  \Gamma^{P}_{\alpha \mu \lambda} \Gamma_{\lambda \nu \alpha}+
    \Gamma_{\alpha \mu \lambda} \Gamma^{P}_{\lambda \nu \alpha} &=&
         4 P_{\mu \nu}(q) d^{-1}(q) \nonumber \\
            & & - 2P_{\mu \nu}(k) d^{-1}(k)   \nonumber \\
            & & - 2P_{\mu \nu}(k+q) d^{-1}(k+q).
\label{Gamc}
\eea
We drop the first term on the RHS of (\ref{Gamc}) following 
the intrinsic PT rule. Now we use the dimensional regularization rule
(which we adhere to throughout this paper)
\be
     \int d^{D} k k^{-2} = 0,
\label{int}
\ee
and discard the parts which disappear after integration, then the second and 
third terms can be written as
\be             
  - 2P_{\mu \nu}(k) d^{-1}(k) - 2P_{\mu \nu}(k+q) d^{-1}(k+q)
           = -2k_{\mu}k_{\nu}-2(k+q)_{\mu}(k+q)_{\nu}.
\label{Gamd}
\ee
Also applying the dimensional regularization rule Eq.(\ref{int}) to the 
fourth term on the RHS of Eq.(\ref{Gamb}), we find
\be
 -\Gamma^{P}_{\alpha \mu \lambda} \Gamma^{P}_{\lambda \nu \alpha}
       = -2k_{\mu}k_{\nu}-(k_{\mu}q_{\nu}+q_{\mu}k_{\nu}).
\label{GamPP}
\ee
Now combining the first term on the RHS of Eq.(\ref{Gamb}) 
with Eqs.(\ref{Gamd}) and (\ref{GamPP}), and inserting them into 
Eq.(\ref{SelfE}), we arrive at  
the following expression for the gauge-invariant self-energy~\cite{rCP}
\bea
     \widehat{\Pi}_{\mu \nu} &=& \frac{iNg^{2}}{2} \int 
               \frac{d^D k}{(2\pi)^D}\frac{1}{k^2 (k+q)^2} \nonumber \\
      & &\qquad \times[\Gamma^{F}_{\alpha \mu \lambda}(k,q) 
                    \Gamma^{F}_{\lambda \nu \alpha}(k+q,-q) 
           -2 (2k+q)_{\mu}(2k+q)_{\nu}].
\label{SelfEb}
\eea

The same rules are applied to obtain the gauge-invariant three-gluon 
vertex at one-loop level.  The contributions of the graphs depicted in 
Fig.3 to the ordinary proper three-gluon vertex are summarized 
as~\cite{defGamma}
\bea
    \Gamma^{0}_{\mu \nu \alpha} &=& -\frac{iNg^{2}}{2} \int 
               \frac{d^D k}{(2\pi)^D}
      \frac{1}{k^2_1 k^2_2 k^2_3}N_{\mu \nu \alpha} \\
    N_{\mu \nu \alpha} &=& \Gamma_{\sigma \mu \lambda}(k_2, q_1)
                           \Gamma_{\lambda \nu \rho}(k_3, q_2)
                \Gamma_{\rho \alpha \sigma}(k_1, q_3) \nonumber \\
          & & \qquad \qquad
           +k_{1\nu}k_{2\alpha}k_{3\mu}+k_{1\alpha}k_{2\mu}k_{3\nu}.
\label{Vert}
\eea
where the momenta and Lorentz indices are defined in Fig.3(a) and 
the overall group-theoretic factor $gf^{abc}$ is omitted.

Decomposing $\Gamma$ into $\Gamma^{F}+\Gamma^{P}$ and dropping the terms 
involving $d^{-1}(q_i)$ which are generated by application of $\Gamma^{P}$ to 
the full vertices, Cornwall and Papavassiliou obtained the following 
expression for gauge-invariant proper three-gluon vertex:
\bea
    & &\widehat{\Gamma}_{\mu \nu \alpha}(q_1, q_2, q_3) =  \nonumber \\
       &-& \frac{iNg^{2}}{2} \biggl\{\int 
           \frac{d^D k}{(2\pi)^D}\frac{1}{k^2_1 k^2_2 k^2_3}
        \biggl[\Gamma^F_{\sigma \mu \lambda}(k_2, q_1)
                   \Gamma^F_{\lambda \nu \rho}(k_3, q_2)
                \Gamma^F_{\rho \alpha \sigma}(k_1, q_3) 
                       \nonumber \\
    & &  \qquad \qquad \qquad \qquad \qquad \qquad \qquad \qquad
        + 2 (k_2+k_3)_{\mu}(k_3+k_1)_{\nu}(k_1+k_2)_{\alpha} 
           \biggr]  \nonumber \\
   & &\qquad \qquad \qquad -8 (q_{1\alpha}g_{\mu \nu}- q_{1\nu}g_{\mu \alpha}) 
    \widetilde{A}(q_{1})
       -8 (q_{2\mu}g_{\alpha \nu}- q_{2\alpha}g_{\mu \nu})\widetilde{A}(q_{2})
          \nonumber \\
   & &  \qquad \qquad \qquad \qquad \qquad \qquad 
   \qquad \qquad \quad -8 (q_{3\nu}g_{\mu \alpha}- 
                      q_{3\mu}g_{\nu \alpha}) \widetilde{A}(q_{3}) \biggr\},
\label{Verta}
\eea
where $\widetilde{A}(q_{i})$ is defined by
\be
     \widetilde{A}(q_{i})= \int \frac{d^D k}{(2\pi)^D}
             \frac{1}{k^2 ( k + q_i )^2}.
\label{Ai}
\ee
\bigskip

\section{The background field method 
         and the gauge-invariant gluon self-energy}
\smallskip

In this section we write down the Feynman 
rules for QCD in the background field calculations and compute 
the gluon self-energy in the Feynman gauge. Then we will see that the result 
coincides, term by term, with the gauge-invariant gluon self-energy 
derived via the intrinsic PT.

In BFM, the field in the classical lagrangian is written as $A+Q$, 
where $A$ ($Q$) denotes the background (quantum) field.
The Feynman diagrams with $A$ on external legs and $Q$ inside loops
need to be calculated.
The relevant Feynman rules~\cite{Abb} are given in Fig.4.
It is noted that the Feynman rule for  
the ghost-$A$ vertex is similar to the one which appears in the scalar QED.
Now let us write a three-point vertex with one $A^b_{\mu}$ field as 
\bea
     \widetilde{\Gamma}^{abc}_{\lambda \mu \nu } (p, q, r) &=& g f^{bac} 
              \widetilde{\Gamma}_{\lambda \mu \nu}(p, q, r)  \\
  \widetilde{\Gamma}_{\lambda \mu \nu}(p, q, r) &=& 
         (p-q+\frac{1}{\xi}r)_{\nu} 
            g_{\lambda\mu} + (q-r-\frac{1}{\xi}p)_{\lambda} g_{\mu\nu} 
                            \nonumber \\
      & &\qquad \qquad \qquad \quad + (r-p)_{\mu} g_{\nu \lambda}  .
\label{vertex}
\eea
Then we find that in the Feynman gauge $\xi=1$,   
$\widetilde{\Gamma}_{\alpha \mu \lambda}(k, q, -k-q)$ turns out to be  
\be
   \widetilde{\Gamma}_{\alpha \mu \lambda}(k, q, -k-q) \vert _{\xi=1}
        = -2 q_{\lambda}g_{\alpha\mu} + 2 q_{\alpha}g_{\mu\lambda} 
              - (2k +q)_{\mu} g_{\lambda \alpha},
\label{verFeyn} 
\ee
which coincides with the expression of $\Gamma^F_{\alpha \mu \lambda}$ in 
Eq.(\ref{FGGG}).   
This fact gives us a hint that BFM may reproduce the same results which 
are obtained by the intrinsic PT (and we find later that it is true in fact).

Now we calculate the gluon self-energy in BFM with the Feynman gauge.
The relevant diagrams are depicted in Fig.5. The diagram 5(a) gives
a contribution 
\be
   \widehat{\Pi}^{(a)}_{\mu \nu} = \frac{iNg^{2}}{2} \int 
               \frac{d^D k}{(2\pi)^D}\frac{1}{k^2 (k+q)^2} 
               \Gamma^{F}_{\alpha \mu \lambda}(k,q) 
                    \Gamma^{F}_{\lambda \nu \alpha}(k+q,-q),
\ee
where we have used the fact 
$\widetilde{\Gamma}_{\alpha \mu \lambda}\vert _{\xi=1} = 
\Gamma^{F}_{\alpha \mu \lambda}$ and 
$\widetilde{\Gamma}_{\lambda \nu \alpha}\vert _{\xi=1} = 
\Gamma^{F}_{\lambda \nu \alpha}$. On the other hand, through the scalar
QED-like 
coupling for the background field and ghost vertices,  
the diagram 5(b) gives 
\be
     \widehat{\Pi}^{(b)}_{\mu \nu} = \frac{iNg^{2}}{2} \int 
               \frac{d^D k}{(2\pi)^D}\frac{1}{k^2 (k+q)^2} 
               [-2 (2k+q)_{\mu}(2k+q)_{\nu}].
\ee
It is interesting to note that the contributions of diagrams 5(a) and 5(b)
, respectively, correspond to 
the first and second terms in the parentheses of Eq.(\ref{SelfEb}) 
and the sum of the two contributions  
coincides with the expression of the gauge-invariant self-energy 
$\widehat{\Pi}_{\mu \nu}$ which was derived in section 2 by the method of 
intrinsic PT.

\bigskip

\section{The gauge-invariant three-gluon vertex}

The success in deriving the gauge-invariant PT result for the gluon 
self-energy by BFM gives us momentum to study, for the next step, 
the gauge-invariant three-gluon vertex at one-loop level. 
The relevant diagrams are 
shown in Fig.6, where momenta and Lorentz and color indices are displayed. 
With the fact that an $AQQ$ vertex in the Feynman gauge,  
$\widetilde{\Gamma}_{\xi=1}$, is equivalent to $\Gamma^F$ in 
Eq.(\ref{FGGG}), it is easy to show that the contribution of the diagram 6(a) 
is
\bea
   ^{(a)}\Gamma^{abc}_{\mu \nu \alpha}(q_1, q_2, q_3) &=&
    - \frac{iNg^{3}f^{abc}}{2} \int 
           \frac{d^D k}{(2\pi)^D}\frac{1}{k^2_1 k^2_2 k^2_3} \nonumber \\
      & & \qquad \qquad \times \biggl[\Gamma^F_{\sigma \mu \lambda}(k_2, q_1) 
         \Gamma^F_{\lambda \nu \rho}(k_3, q_2)
                \Gamma^F_{\rho \alpha \sigma}(k_1, q_3) \biggr].
\eea
The contribution of the diagram 6(b) (and the similar one with the ghost 
running the other way) is 
\bea
 ^{(b)}\Gamma^{abc}_{\mu \nu \alpha}(q_1, q_2, q_3) &=&
    - \frac{iNg^{3}f^{abc}}{2} \int 
           \frac{d^D k}{(2\pi)^D}\frac{1}{k^2_1 k^2_2 k^2_3} \nonumber \\
  & & \qquad \qquad \times \biggl[2 (k_2+k_3)_{\mu}(k_3+k_1)_{\nu}
         (k_1+k_2)_{\alpha} \biggr].
\eea
When we calculate the diagram 6(c), again we use the Feynman gauge ($\xi=1$) 
for the four-point vertex with two background fields. Remembering that the 
diagram 6(c) has a symmetric factor $\frac{1}{2}$ and adding 
the two other similar diagrams, we find 
\bea
^{(c)}\Gamma^{abc}_{\mu \nu \alpha}(q_1, q_2, q_3) &=&
    \frac{iNg^{3}f^{abc}}{2}
     \biggl[8 (q_{1\alpha}g_{\mu \nu}- q_{1\nu}g_{\mu \alpha}) 
     \widetilde{A}(q_{1})     \nonumber \\
    & & +8 (q_{2\mu}g_{\alpha \nu}- q_{2\alpha}g_{\mu \nu}) 
    \widetilde{A}(q_{2}) 
   +8 (q_{3\nu}g_{\mu \alpha}- q_{3\mu}g_{\nu \alpha})
                  \widetilde{A}(q_{3}) \biggr].
\eea
Finally, the contribution of the diagram 6(d) (and two other similar diagrams) 
turns out to be null because of the group-theoretical identity 
for the structure constants $f^{abc}$ such as
\be
    f^{ead} ( f^{dbx} f^{xce}+f^{dcx}f^{xbe}) = 0.
\ee

Now adding the contributions from the diagrams (a)-(c) in Fig. 6 
and omitting the overall group-theoretic factor $gf^{abc}$, we find that 
the result coincides with the expression of Eq.(\ref{Verta}) which was obtained 
by the intrinsic PT. Also we note that each contribution from 
the diagrams (a)-(c), respectively, 
corresponds to a particular term in Eq.(\ref{Verta}).

Finally we close this section with a mention that the constructed 
$\widehat{\Gamma}_{\mu \nu \alpha}(q_1, q_2, q_3)$ 
is related to the gauge-invariant self-energy $\widehat{\Pi}_{\mu \nu}$ 
of Eq.(\ref{SelfEb}) through a Ward identity~\cite{rCP} 
\be
   q_1^{\mu}\widehat{\Gamma}_{\mu \nu \alpha}(q_1, q_2, q_3) = 
               -\widehat{\Pi}_{\nu \alpha}(q_2)  
                + \widehat{\Pi}_{\nu \alpha}(q_3),
\ee    
which is indeed a naive extension of the tree-level one.

\bigskip
\section{The gauge-invariant four-gluon vertex and its Ward identity}
\smallskip

The gauge-invariant 
four-gluon vertex has been constructed by Papavassiliou~\cite{rPa} 
using the $S$-matrix PT. As he stated in Ref.~\cite{rPa}, the 
construction was a nontrivial task because of the large number of graphs 
and certain subtleties of PT. Although he did not report 
the exact closed form of the gauge-invariant four-gluon vertex, he showed that 
the new four-gluon vertex is related to the previously constructed 
$\widehat{\Gamma}_{\mu \nu \alpha}$ in Eq.(\ref{Verta}) through 
a simple Ward identity. 
In this section we apply BFM with the Feynman gauge to obtain 
the gauge-invariant four-gluon vertex at one-loop level. We give the closed 
form of this vertex and show that it satisfies the same Ward identity which 
was proved by Papavassiliou.

The bare four-gluon vertex in Fig.7(a) is expressed as 
$-ig^2\Gamma^{abcd}_{\mu\nu\alpha\beta}$ with 
\bea
 \Gamma^{abcd}_{\mu\nu\alpha\beta}&=&
        f^{abx} f^{cdx}(g_{\mu\alpha}g_{\nu\beta}
                            -g_{\mu\beta}g_{\nu\alpha}) \nonumber \\
    &+& f^{acx} f^{dbx}(g_{\mu\beta}g_{\alpha\nu}
                            -g_{\mu\nu}g_{\alpha\beta}) \nonumber \\
    &+& f^{adx} f^{bcx}(g_{\mu\nu}g_{\beta\alpha}
                            -g_{\mu\alpha}g_{\beta\nu}),
\eea
while the bare three-gluon vertex in Fig.7(b) is expressed as 
$g\Gamma^{abc}_{\mu\nu\lambda}(k_1,k_2,k_3)$ with 
\be
   \Gamma^{abc}_{\mu\nu\lambda}(k_1,k_2,k_3)=
          f^{abc}\Gamma_{\mu\nu\lambda}(k_1,k_2,k_3)
\ee
and $\Gamma_{\mu\nu\lambda}(k_1,k_2,k_3)$ is given by Eq.(\ref{Gam}). 
Now acting with $q^{\mu}_1$ on $\Gamma^{abcd}_{\mu\nu\alpha\beta}$, 
we get
\bea
 q^{\mu}_1\Gamma^{abcd}_{\mu\nu\alpha\beta}&=&
        f^{abx} f^{cdx}(q_{1\alpha}g_{\nu\beta}
                            -q_{1\beta}g_{\nu\alpha}) \nonumber \\
    &+& f^{acx} f^{dbx}(q_{1\beta}g_{\alpha\nu}
                            -q_{1\nu}g_{\alpha\beta}) \nonumber \\
    &+& f^{adx} f^{bcx}(q_{1\nu}g_{\beta\alpha}
                            -q_{1\alpha}g_{\beta\nu}).
\label{BWa}
\eea 
Next with a help of the Jacobi identity
\be
    f^{abx} f^{cdx}+f^{acx} f^{dbx}+f^{adx} f^{bcx} = 0, 
\label{Jacobi}
\ee
we add
\bea
   0 &=& \biggl( f^{abx} f^{cdx}+f^{acx} f^{dbx}+f^{adx} f^{bcx} \biggr) 
                   \nonumber \\
     & & \times \biggl[ (q_{4}-q_{3})_{\nu}g_{\alpha\beta}
           + (q_{2}-q_{4})_{\alpha}g_{\beta\nu} 
           + (q_{3}-q_{2})_{\beta}g_{\nu\alpha} \biggr]
\eea
to the RHS of Eq.(\ref{BWa}) and we obtain the tree-level Ward 
identity~\cite{defGamma}
\bea
 q^{\mu}_1\Gamma^{abcd}_{\mu\nu\alpha\beta}&=&
        -f^{abx} \Gamma^{cdx}_{\alpha\beta\nu}(q_3,q_4,q_1+q_2)
                             \nonumber \\
    & & -f^{acx} \Gamma^{dbx}_{\beta\nu\alpha}(q_4,q_2,q_1+q_3)
                             \nonumber \\
    & & -f^{adx} \Gamma^{bcx}_{\nu\alpha\beta}(q_2,q_3,q_1+q_4).
\label{BWard}
\eea 
The bare three- and four-gluon vertices are manifestly gauge independent. 
However, if we consider the {\sl usual} one-loop corrections to these vertices, 
they become gauge dependent and do not satisfy Eq.(\ref{BWard}) any more. 

We now apply BFM to the case of the four-gluon vertex and show that
the constructed gauge-invariant vertex satisfies the generalized version of 
the Ward identity in Eq.(\ref{BWard}). The diagrams for the four-gluon vertex 
at one-loop level are shown in Fig.8.  It is noted that 
the two-gluon loop diagrams 8(e) 
have a symmetric factor $\frac{1}{2}$. For later convenience let us
introduce the following group-theoretic quantities:
\be
f(abcd) \equiv f^{alm} f^{bmn} f^{cne} f^{del},
\ee
which satisfy the relations
\bea
f(abcd)=f(bcda)=f(badc) \label{cycle} \\
f(abcd)-f(abdc)=-\frac{N}{2}f^{abx}f^{cdx}.
\label{frel} 
\eea
The last relation is derived from 
an identity for the structure constants $f^{abc}$
\be
f^{alm}f^{bmn}f^{cnl}=\frac{N}{2}f^{abc}
\label{fiden}
\ee
and the Jacobi identity Eq.(\ref{Jacobi}). 

It is straightforward to evaluate the diagrams in Fig.8, and  
the relevant momenta, color and Lorentz indices are indicated in the graphs. 
The relations for color factors in Eqs.(\ref{cycle})-(\ref{frel}) are 
extensively used.  We present each contribution to the vertex from the 
individual Feynman diagram, expecting that each contribution 
corresponds to the particular term of the intrinsic PT result once 
the calculation is made in the future. 
The results are the following. 
\medskip

\noindent
[a]: The diagrams 8(a) give
\bea
& &^{(a)}\widehat{\Gamma}^{abcd}_{\mu \nu \alpha \beta}(q_1, q_2, q_3, q_4) =
   ig^2 f(abcd)\int \frac{d^{D}k}{(2\pi)^D}
       \frac{1}{k^2 (k+q_2)^2 (k-q_1-q_4)^2 (k-q_1)^2} \nonumber \\
 & & \qquad \qquad 
            \times \biggl[\Gamma^F_{\lambda \mu \rho}(k-q_1, q_1)
            \Gamma^F_{\rho \nu \tau}(k, q_2)
            \Gamma^F_{\tau \alpha \kappa}(k+q_2, q_3)
            \Gamma^F_{\kappa \beta \lambda}(k-q_1-q_4, q_4)\biggr] \nonumber \\
 & & \qquad \qquad 
     + \biggl\{ (q_2, b, \nu) \longleftrightarrow (q_3, c, \alpha) \biggl\}
     + \biggl\{ (q_3, c, \alpha) \longleftrightarrow (q_4, d, \beta) \biggl\},
\label{fourgluona}
\eea
where the notation 
$\biggl\{ (q_2, b, \nu) \longleftrightarrow (q_3, c, \alpha) \biggl\}$ 
represents a term obtained from the first one by the substitution 
$(q_2, b, \nu) \longleftrightarrow (q_3, c, \alpha)$. The same notation 
applies to the third term, and also to the expressions below.

\noindent
[b]: The diagrams 8(b) give
\bea
& &^{(b)}\widehat{\Gamma}^{abcd}_{\mu \nu \alpha \beta}(q_1, q_2, q_3, q_4) =
   -ig^2 f(abcd)\int \frac{d^{D}k}{(2\pi)^D}
       \frac{1}{k^2 (k+q_2)^2 (k-q_1-q_4)^2 (k-q_1)^2} \nonumber \\
 & & \qquad \qquad \qquad
       \times \biggl[2(2k-q_1)_{\mu}(2k+q_2)_{\nu}
                (2k-q_1-q_4+q_2)_{\alpha}(2k-2q_1-q_4)_{\beta}
            \biggr] \nonumber \\
 & & \qquad \qquad \qquad
     + \biggl\{ (q_2, b, \nu) \longleftrightarrow (q_3, c, \alpha) \biggl\}
     + \biggl\{ (q_3, c, \alpha) \longleftrightarrow (q_4, d, \beta) \biggl\}.
\eea

\noindent
[c]: The diagrams 8(c) give
\bea
& &^{(c)}\widehat{\Gamma}^{abcd}_{\mu \nu \alpha \beta}(q_1, q_2, q_3, q_4) =
   ig^2 \int \frac{d^{D}k}{(2\pi)^D}
       \frac{1}{k^2 (k+q_2)^2 (k-q_1)^2} \nonumber \\
 & & \qquad 
        \times \biggl\{-\biggl[ f(abcd)+f(abdc) \biggr] g_{\alpha\beta}
           \Gamma^F_{\lambda \mu \rho}(k-q_1, q_1)
            \Gamma^F_{\rho \nu \lambda}(k, q_2)  \nonumber \\
 & & \qquad \quad + Nf^{abx}f^{cdx}\biggl[
          \Gamma^F_{\beta \mu \lambda}(k-q_1, q_1)
            \Gamma^F_{\lambda \nu \alpha}(k, q_2) 
           - \Gamma^F_{\alpha \mu \lambda}(k-q_1, q_1)
            \Gamma^F_{\lambda \nu \beta}(k, q_2) \biggr] \biggr\}    
                   \nonumber \\
 & & \qquad 
     + \biggl\{ (q_2, b, \nu) \longleftrightarrow (q_3, c, \alpha) \biggl\}
     + \biggl\{ (q_2, b, \nu) \longleftrightarrow (q_4, d, \beta) \biggl\}
                  \nonumber \\
 & & \qquad 
     + \biggl\{ (q_1, a, \mu) \longleftrightarrow (q_3, c, \alpha) \biggl\}
     + \biggl\{ (q_1, a, \mu) \longleftrightarrow (q_4, d, \beta) \biggl\}
                    \nonumber \\
 & & \qquad 
     + \biggl\{ (q_1, a, \mu) \longleftrightarrow (q_3, c, \alpha), \quad 
                 (q_2, b, \nu) \longleftrightarrow (q_4, d, \beta) \biggl\}.
\eea

\noindent
[d]: The diagrams 8(d) give
\bea
& &^{(d)}\widehat{\Gamma}^{abcd}_{\mu \nu \alpha \beta}(q_1, q_2, q_3, q_4) =
   ig^2 \int \frac{d^{D}k}{(2\pi)^D}
       \frac{1}{k^2 (k+q_2)^2 (k-q_1)^2} \nonumber \\
 & & \qquad 
        \times \biggl[ f(abcd)+f(abdc) \biggr] 2 g_{\alpha\beta}
           (2k-q_1)_{\mu} (2k+q_2)_{\nu} \nonumber \\
 & & \qquad 
     + \biggl\{ (q_2, b, \nu) \longleftrightarrow (q_3, c, \alpha) \biggl\}
     + \biggl\{ (q_2, b, \nu) \longleftrightarrow (q_4, d, \beta) \biggl\}
                  \nonumber \\
 & & \qquad 
     + \biggl\{ (q_1, a, \mu) \longleftrightarrow (q_3, c, \alpha) \biggl\}
     + \biggl\{ (q_1, a, \mu) \longleftrightarrow (q_4, d, \beta) \biggl\}
                    \nonumber \\
 & & \qquad 
     + \biggl\{ (q_1, a, \mu) \longleftrightarrow (q_3, c, \alpha), \quad 
                 (q_2, b, \nu) \longleftrightarrow (q_4, d, \beta) \biggl\}.
\eea

\noindent
[e]: The diagrams 8(e) give
\bea
& &^{(e)}\widehat{\Gamma}^{abcd}_{\mu \nu \alpha \beta}(q_1, q_2, q_3, q_4) =
    ig^2 \widetilde{A}(q_1+q_2) \nonumber \\
  & & \qquad \qquad \times \biggl\{ \biggl[f(abcd)+f(abdc)\biggr] 
               D g_{\mu\nu}g_{\alpha\beta} + 
    4Nf^{abx}f^{cdx}(g_{\mu\alpha}g_{\nu\beta}
                            -g_{\mu\beta}g_{\nu\alpha})\biggl\}
                      \nonumber \\
 & & \qquad \qquad
     +  \biggl\{(q_2, b, \nu) \longleftrightarrow (q_3, c, \alpha) \biggl\}
     + \biggl\{ (q_2, b, \nu) \longleftrightarrow (q_4, d, \beta) \biggl\}.
\eea
where $\widetilde{A}(q_{i})$ is defined in Eq.(\ref{Ai}).

\noindent
[f]: The diagrams 8(f) give
\bea
& &^{(f)}\widehat{\Gamma}^{abcd}_{\mu \nu \alpha \beta}(q_1, q_2, q_3, q_4) =
    -ig^2 \widetilde{A}(q_1+q_2) \biggl[f(abcd)+f(abdc)\biggr] 
                    2 g_{\mu\nu}g_{\alpha\beta}
                      \nonumber \\
 & & \qquad \qquad
     +  \biggl\{(q_2, b, \nu) \longleftrightarrow (q_3, c, \alpha) \biggl\}
     + \biggl\{ (q_2, b, \nu) \longleftrightarrow (q_4, d, \beta) \biggl\}.
\label{ffgluon}
\eea

Then the final form of the gauge-invariant four-gluon vertex 
$\widehat{\Gamma}^{abcd}_{\mu \nu \alpha \beta}$ at one-loop level 
is given by the sum
\bea
   \widehat{\Gamma}^{abcd}_{\mu \nu \alpha \beta} &=&
         ^{(a)}\widehat{\Gamma}^{abcd}_{\mu \nu \alpha \beta} +
         ^{(b)}\widehat{\Gamma}^{abcd}_{\mu \nu \alpha \beta} +
         ^{(c)} \widehat{\Gamma}^{abcd}_{\mu \nu \alpha \beta} 
                    \nonumber \\
         &+& ^{(d)}\widehat{\Gamma}^{abcd}_{\mu \nu \alpha \beta}
          + ^{(e)}\widehat{\Gamma}^{abcd}_{\mu \nu \alpha \beta}
          + ^{(f)}\widehat{\Gamma}^{abcd}_{\mu \nu \alpha \beta}
\label{fourVertex}
\eea

Our next task is to show explicitly that the above four-gluon vertex 
$\widehat{\Gamma}^{abcd}_{\mu \nu \alpha \beta}$ satisfies the 
following generalized 
version of the Ward Identity in Eq.(\ref{BWard}):
\bea
 q^{\mu}_1\widehat{\Gamma}^{abcd}_{\mu\nu\alpha\beta}&=&
        -f^{abx} \widehat{\Gamma}^{cdx}_{\alpha\beta\nu}(q_3,q_4,q_1+q_2)
                             \nonumber \\
    & & -f^{acx} \widehat{\Gamma}^{dbx}_{\beta\nu\alpha}(q_4,q_2,q_1+q_3)
                             \nonumber \\
    & & -f^{adx} \widehat{\Gamma}^{bcx}_{\nu\alpha\beta}(q_2,q_3,q_1+q_4),
\label{WardIden}
\eea 
where 
\be
   \widehat{\Gamma}^{abc}_{\mu\nu\lambda}(q_1,q_2,q_3) 
       =f^{abc}\widehat{\Gamma}_{\mu\nu\lambda}(q_1,q_2,q_3)
\ee
and $\widehat{\Gamma}_{\mu\nu\lambda}(q_1,q_2,q_3)$ is the gauge-invariant 
three-gluon vertex given in Eq.(\ref{Verta}). 
The Ward identity Eq.(43), first found by Papavassiliou with PT [4], 
is naturally expected to hold in BFM formalism.

We act with $q_1^{\mu}$ on the individual contributions to 
$\widehat{\Gamma}^{abcd}_{\mu\nu\alpha\beta}$ which are expressed in 
Eqs.(\ref{fourgluona})-(\ref{ffgluon}). Before going through 
the evaluation we make some 
preparations. Let us introduce the following integrals 
for the three-point vertex 
with the constraint $q_1 + q_2 + q_3 = 0$:
\bea
 & & B_{\mu\nu\alpha}(q_1,q_2,q_3) \equiv \int 
   \frac{d^D k}{(2\pi)^D}\frac{1}{k^2 (k+q_2)^2 (k-q_1)^2} \nonumber \\
 & & \qquad \qquad \qquad \qquad \qquad 
      \times \biggl[\Gamma^F_{\lambda \mu \rho}(k-q_1, q_1)
            \Gamma^F_{\rho \nu \tau}(k, q_2)
            \Gamma^F_{\tau \alpha \lambda}(k+q_2, q_3) \biggr], \nonumber \\
 & & C_{\mu\nu\alpha}(q_1,q_2,q_3) \equiv \int 
   \frac{d^D k}{(2\pi)^D}\frac{1}{k^2 (k+q_2)^2 (k-q_1)^2} \nonumber \\
    & & \qquad \qquad \qquad \qquad \qquad 
     \times (2k-q_1)_{\mu}(2k+q_2)_{\nu}
                (2k-q_1+q_2)_{\alpha}.
\eea
In terms of $B_{\mu\nu\alpha}$ and $C_{\mu\nu\alpha}$, 
the gauge-invariant three-gluon vertex $\widehat{\Gamma}_{\mu\nu\lambda}$ in 
Eq.(\ref{Verta}) is expressed as 
\bea
\widehat{\Gamma}_{\mu\nu\lambda}(q_1,q_2,q_3) &=&-\frac{iNg^2}{2}
   \biggl\{ B_{\mu\nu\alpha}(q_1,q_2,q_3) + 2 C_{\mu\nu\alpha}(q_1,q_2,q_3) 
            \nonumber \\
  & & \quad  -8 (q_{1\alpha}g_{\mu \nu}- q_{1\nu}g_{\mu \alpha}) 
    \widetilde{A}(q_{1})
       -8 (q_{2\mu}g_{\alpha \nu}- q_{2\alpha}g_{\mu \nu})\widetilde{A}(q_{2})
          \nonumber \\
& &  \qquad \qquad \qquad \qquad 
   \qquad  -8 (q_{3\nu}g_{\mu \alpha}- 
                      q_{3\mu}g_{\nu \alpha}) \widetilde{A}(q_{3}) \biggr\}.
\label{newDefG}
\eea
These $B_{\mu\nu\alpha}$ and $C_{\mu\nu\alpha}$ satisfy the following 
relations:
\bea
  B_{\mu\nu\alpha}(q_1,q_2,q_3) &=& B_{\nu\alpha\mu}(q_2,q_3,q_1) = 
                 -B_{\nu\mu\alpha}(q_2,q_1,q_3),  \label{BBB} \\
  C_{\mu\nu\alpha}(q_1,q_2,q_3) &=& C_{\nu\alpha\mu}(q_2,q_3,q_1) = 
                 -C_{\nu\mu\alpha}(q_2,q_1,q_3), \label{CCC}
\eea
which can be proved by changing integration variables under the 
constraint $q_1 + q_2 + q_3 = 0$ and using the fact 
\be
   \Gamma^F_{\lambda \mu \rho}(k, q) = 
              - \Gamma^F_{\rho \mu \lambda}(-k-q, q).
\label{tranGamma}
\ee

Throughout the algebraic manipulations, we often take the means of changing the 
integration variables under the constraint 
$q_1+q_2+q_3+q_4=0$ and make use of  
identities
\bea
    q_1^{\mu}\Gamma^{F}_{\lambda\mu\rho}(k-q_1,q_1)
            &=&\biggl[(k-q_1)^2-k^2\biggr] g_{\lambda\rho}, \nonumber \\
    q_1^{\mu}(2k-q_1)_{\mu} &=& k^2-(k-q_1)^2
\eea
to reduce the number of propagators by one, the relations of 
Eqs.(\ref{BBB})-(\ref{tranGamma}) for $B_{\mu\nu\alpha}$, $C_{\mu\nu\alpha}$ 
and $\Gamma^F_{\lambda \mu \rho}$ to classify terms into groups with 
the same color factors. We also use the identity of Eq.(\ref{frel}). 
The results are 

\medskip

\noindent
[a]:
\bea
 & & q^{\mu}_1\ ^{(a)}
\widehat{\Gamma}^{abcd}_{\mu \nu \alpha \beta}(q_1, q_2, q_3, q_4) =
     \nonumber \\
 & & \qquad ig^2 \biggl\{\biggl[ f(abcd) + f(abdc) \biggr] q_{1\nu} 
    \int \frac{d^D k}{(2\pi)^D}
         \frac{\Gamma^F_{\lambda \alpha \rho}(k-q_3, q_3)
    \Gamma^F_{\rho \beta \lambda}(k, q_4)}{(k-3)^2 k^2 (k+q_4)^2}
               \nonumber \\
 & & \qquad \qquad + N f^{abx}f^{cdx}
        \biggl[\frac{1}{2}B_{\alpha\beta\nu}(q_3,q_4,q_1+q_2) \nonumber \\
 & & \quad \qquad \qquad \quad +q_1^{\lambda}\int \frac{d^D k}{(2\pi)^D}
         \frac{\Gamma^F_{\lambda \alpha \rho}(k-q_3, q_3)
    \Gamma^F_{\rho \beta \nu}(k, q_4) -
   (\nu \longleftrightarrow \lambda)}{(k-3)^2 k^2 (k+q_4)^2}\biggr] \biggr\}
   \nonumber \\
& & \qquad \quad + \biggl\{ \rm{cyclic \ permutations} \biggl\}
\label{qgamma}
\eea
where $\bigl\{ \rm{cyclic \ permutations} \bigl\}$ represents 
two terms which are obtained from the first term by the 
substitution 
$\bigl\{ (q_2,b,\nu) \rightarrow (q_3,c,\alpha) \rightarrow (q_4,d,\beta) 
\rightarrow (q_2,b,\nu) \bigr\}$ and the substitution 
$\bigl\{ (q_2,b,\nu) \rightarrow (q_4,d,\beta) \rightarrow (q_3,c,\alpha) 
\rightarrow (q_2,b,\nu) \bigr\}$. The same notation applies to the 
expressions below.

\noindent
[b]:
\bea
 & & q^{\mu}_1\ ^{(b)}
\widehat{\Gamma}^{abcd}_{\mu \nu \alpha \beta}(q_1, q_2, q_3, q_4) =
     \nonumber \\
 & & \qquad ig^2 \biggl\{\biggl[ f(abcd) + f(abdc) \biggr](-2 q_{1\nu}) 
   \int \frac{d^D k}{(2\pi)^D}
         \frac{(2k-q_3)_{\alpha}(2k+q_4)_{\beta}}{(k-3)^2 k^2 (k+q_4)^2}
               \nonumber \\
 & & \qquad \qquad + N f^{abx}f^{cdx} C_{\alpha\beta\nu}(q_3,q_4,q_1+q_2) \biggr\}
          \nonumber \\
& &  \qquad \quad + \biggl\{ \rm{cyclic \ permutations} \biggl\}
\label{qgammb}
\eea
\noindent
[c]:
\bea
 & & q^{\mu}_1\ ^{(c)}
\widehat{\Gamma}^{abcd}_{\mu \nu \alpha \beta}(q_1, q_2, q_3, q_4) =
     \nonumber \\
 & & \qquad ig^2 \Biggl( \biggl[ f(abcd) + f(abdc) \biggr] \biggl\{ 
           -D q_{1\nu}g_{\alpha\beta} \widetilde{A}(q_1+q_2) \nonumber \\
 & &  \qquad \qquad \qquad \qquad \qquad \qquad \quad
 -q_{1\nu} \int \frac{d^D k}{(2\pi)^D}
         \frac{\Gamma^F_{\lambda \alpha \rho}(k-q_3, q_3)
    \Gamma^F_{\rho \beta \lambda}(k, q_4)}{(k-3)^2 k^2 (k+q_4)^2} \biggr\}
               \nonumber \\
 & & \qquad \qquad + N f^{abx}f^{cdx} \biggl\{ 
      4(q_{2\alpha}g_{\beta\nu}-q_{2\beta}g_{\nu\alpha})
        \biggl[\widetilde{A}(q_1+q_2)-\widetilde{A}(q_2) \biggr] \nonumber \\
 & & \quad \qquad \qquad \qquad \quad -q_1^{\lambda}\int \frac{d^D k}{(2\pi)^D}
         \frac{\Gamma^F_{\lambda \alpha \rho}(k-q_3, q_3)
    \Gamma^F_{\rho \beta \nu}(k, q_4) -
   (\nu \longleftrightarrow \lambda)}{(k-3)^2 k^2 (k+q_4)^2}\biggr\} \Biggr)
   \nonumber \\
& & \qquad \quad + \biggl\{ \rm{cyclic \ permutations} \biggl\}
\label{qgammc}
\eea
\noindent
[d]:
\bea
 & & q^{\mu}_1\ ^{(d)}
\widehat{\Gamma}^{abcd}_{\mu \nu \alpha \beta}(q_1, q_2, q_3, q_4) =
     \nonumber \\
 & & \qquad ig^2 \biggl[ f(abcd) + f(abdc) \biggr]\biggl\{
         2q_{1\nu}g_{\alpha\beta} \widetilde{A}(q_1+q_2) \nonumber \\
 & &\qquad \qquad \qquad \qquad \qquad \qquad \qquad
       +2 q_{1\nu} \int \frac{d^D k}{(2\pi)^D}
         \frac{(2k-q_3)_{\alpha}(2k+q_4)_{\beta}}{(k-3)^2 k^2 (k+q_4)^2}
          \biggr\}     \nonumber \\
& & \qquad \quad + \biggl\{ \rm{cyclic \ permutations} \biggl\}.
\label{qgammd}
\eea
\noindent
[e]:
\bea
q^{\mu}_1\ ^{(e)}
\widehat{\Gamma}^{abcd}_{\mu \nu \alpha \beta}(q_1, q_2, q_3, q_4) &=& 
 ig^2 \biggl\{ \biggl[ f(abcd) + f(abdc) \biggr] 
           D q_{1\nu}g_{\alpha\beta} \widetilde{A}(q_1+q_2) \nonumber \\
 & & \qquad + N f^{abx}f^{cdx}  
      4(q_{1\alpha}g_{\beta\nu}-q_{1\beta}g_{\nu\alpha})
        \widetilde{A}(q_1+q_2) \biggr\}\nonumber \\
& & \quad + \biggl\{ \rm{cyclic \ permutations} \biggl\}
\label{qgamme}
\eea
\noindent
[f]:
\bea
q^{\mu}_1\ ^{(f)}
\widehat{\Gamma}^{abcd}_{\mu \nu \alpha \beta}(q_1, q_2, q_3, q_4) &=&
   ig^2 \biggl[ f(abcd) + f(abdc) \biggr]
        (-2q_{1\nu})g_{\alpha\beta} \widetilde{A}(q_1+q_2) \nonumber \\
& &  \quad + \biggl\{ \rm{cyclic \ permutations} \biggl\}.
\label{qgammf}
\eea
\medskip

Adding together all the contributions [Eqs.(\ref{qgamma})-(\ref{qgammf})], 
we find 
\bea
q^{\mu}_1\widehat{\Gamma}^{abcd}_{\mu \nu \alpha \beta}(q_1, q_2, q_3, q_4) &=&
   \frac{ig^2 N}{2}f^{abx}f^{cdx} \biggl[ B_{\alpha\beta\nu}(q_3,q_4,q_1+q_2) 
    +2C_{\alpha\beta\nu}(q_3,q_4,q_1+q_2) \nonumber \\
& & \qquad \qquad
    -8\biggl\{(q_1+q_2)_{\beta}g_{\nu\alpha}
                 -(q_1+q_2)_{\alpha}g_{\nu\beta}\biggr\}
        \widetilde{A}(q_1+q_2)   \nonumber \\
& &  \qquad \qquad \qquad  
    + 8\biggl\{q_{2\beta}g_{\nu\alpha}
                 -q_{2\alpha}g_{\nu\beta}\biggr\}
        \widetilde{A}(q_2)    \biggr]    \nonumber \\
& &  \quad + \biggl\{ \rm{cyclic \ permutations} \biggl\}
\label{WIFG}
\eea
It is noted that all the terms which are proportional to factors 
$[f(abcd)+f(abdc)]$, $[f(acdb)+f(acbd)]$, and $[f(adbc)+f(adcb)]$ cancel out 
and only terms with factors $Nf^{abx}f^{cdx}$, $Nf^{acx}f^{dbx}$, 
and $Nf^{adx}f^{bcx}$ remain in the final result. The last step is to add 
\bea
  0 &=& \frac{ig^2 N}{2}
     \biggl[f^{abx}f^{cdx}+f^{acx}f^{dbx}+f^{adx}f^{bcx}\biggr] \nonumber \\
   & & \qquad \times \biggl\{ 
          -8 ( q_{2\beta}g_{\nu\alpha}
                 -q_{2\alpha}g_{\nu\beta}) \widetilde{A}(q_2) \nonumber \\
   & & \qquad \qquad 
        -8 ( q_{3\nu}g_{\alpha\beta}
                 -q_{3\beta}g_{\alpha\nu}) \widetilde{A}(q_3) \nonumber \\
   & & \qquad \qquad 
        -8 ( q_{4\alpha}g_{\beta\nu}
                 -q_{4\nu}g_{\beta\alpha}) \widetilde{A}(q_4) \biggr\}
\eea
to the RHS of Eq.(\ref{WIFG}) and to use Eq.(\ref{newDefG}), and we arrive 
at the desired result of Eq.(\ref{WardIden}).

\bigskip
\section{Conclusions}
\smallskip

In this paper we demonstrated that the background field method is an 
alternative and simple way of deriving the same gauge-invariant results which 
are obtained by the pinch technique. 
We have found, in particular, in the cases of gauge-invariant gluon 
self-energy and three-gluon vertex, that both BFM in the 
Feynman gauge and the intrinsic PT produce the same results which 
are equal term by term. 
We also calculated the gauge-invariant four-gluon vertex in BFM and presented 
its exact form. Finally we explicitly showed that this four-gluon
vertex satisfies the 
same simple Ward identity that was found with PT.

We already know that PT works in 
spontaneously broken gauge theories, especially, in the standard model. 
It will be very interesting to investigate whether BFM may reproduce the 
same results for the one-loop gauge-invariant $WW$ and $ZZ$ self-energies  
and $\gamma WW$ and $ZWW$ vertices which were constructed by PT. 
\bigskip

\vspace{2cm}
\begin{center}
{\large\bf Acknowledgements}
\end{center}
\medskip
One of the authors (K.S) would like to thank Professor D. Zwanziger and 
Joannis Papavassiliou for the hospitality extended to him in the 
spring of 1994 at New York University where part of this work was done, 
and Professor T. Muta for the hospitality extended to him at 
Hiroshima University. He is also very grateful to Joannis Papavassiliou 
for informative and helpful discussions on the pinch technique.


\newpage
\baselineskip 16pt
\noindent
{\large\bf Figure captions}
\medskip

\noindent
Fig.1

\noindent
The $S$-matrix pinch technique applied for the elastic scattering of 
two fermions. Graphs (b) and (c) are pinch parts, which, when added 
to the ordinary propagator graphs (a), yield the gauge-invariant 
effective gluon propagator. 

\medskip

\noindent
Fig.2

\noindent
Graphs for the ordinary proper self-energy $\Pi^{0}_{\mu\nu}$. \  
(a): Gluon-loop.\  (b): Ghost-loop. Momenta and Lorentz indices are 
indicated.

\medskip

\noindent
Fig.3

\noindent
Graphs for the ordinary proper three-gluon vertex 
$\Gamma^{0}_{\mu\nu\lambda}$. \  
(a): Gluon-loop.\ (b)(c): Ghost-loops. 
Momenta and Lorentz indices are indicated.

\medskip

\noindent
Fig.4

\noindent
Feynman rules for background field calculations in QCD. 
The wavy lines terminating in an $A$ represent external gauge fields. 
The other wavy lines and dashed lines represent $Q$ fields and ghost fields, 
respectively. Only shown are rules which are requisite for calculations 
in this paper.

\medskip

\noindent
Fig.5

\noindent
Graphs for a calculation of the gauge-invariant 
self-energy $\widehat{\Pi}_{\mu\nu}$ in BFM. \  
(a):Gluon-loop.\  (b) Ghost-loop. Momenta and Lorentz indices are 
indicated.

\medskip

\noindent
Fig.6

\noindent
Graphs for a calculation of the gauge-invariant 
three-gluon vertex $\widehat{\Gamma}_{\mu\nu\lambda}$ in BFM. \  
(a)(c): Gluon-loops.\  (b)(d): Ghost-loops. 
Momenta and Lorentz indices are indicated.

\medskip

\noindent
Fig.7

\noindent
The bare four-gluon vertex (a) and the bare three-gluon vertex (b). 
Momenta, and color and Lorentz indices are indicated.

\medskip

\noindent
Fig.8

\noindent
Graphs for a calculation of the gauge-invariant 
four-gluon vertex $\widehat{\Gamma}_{\mu\nu\alpha\beta}$ in BFM. \   
(a)(c)(e): Gluon-loops.\ (b)(d)(f): Ghost-loops. 
Momenta, and color and Lorentz indices are indicated.

\input epsf.sty
\pagestyle{empty}


\begin{center}

\hspace{0.2cm}
\epsfxsize=10cm
\epsffile{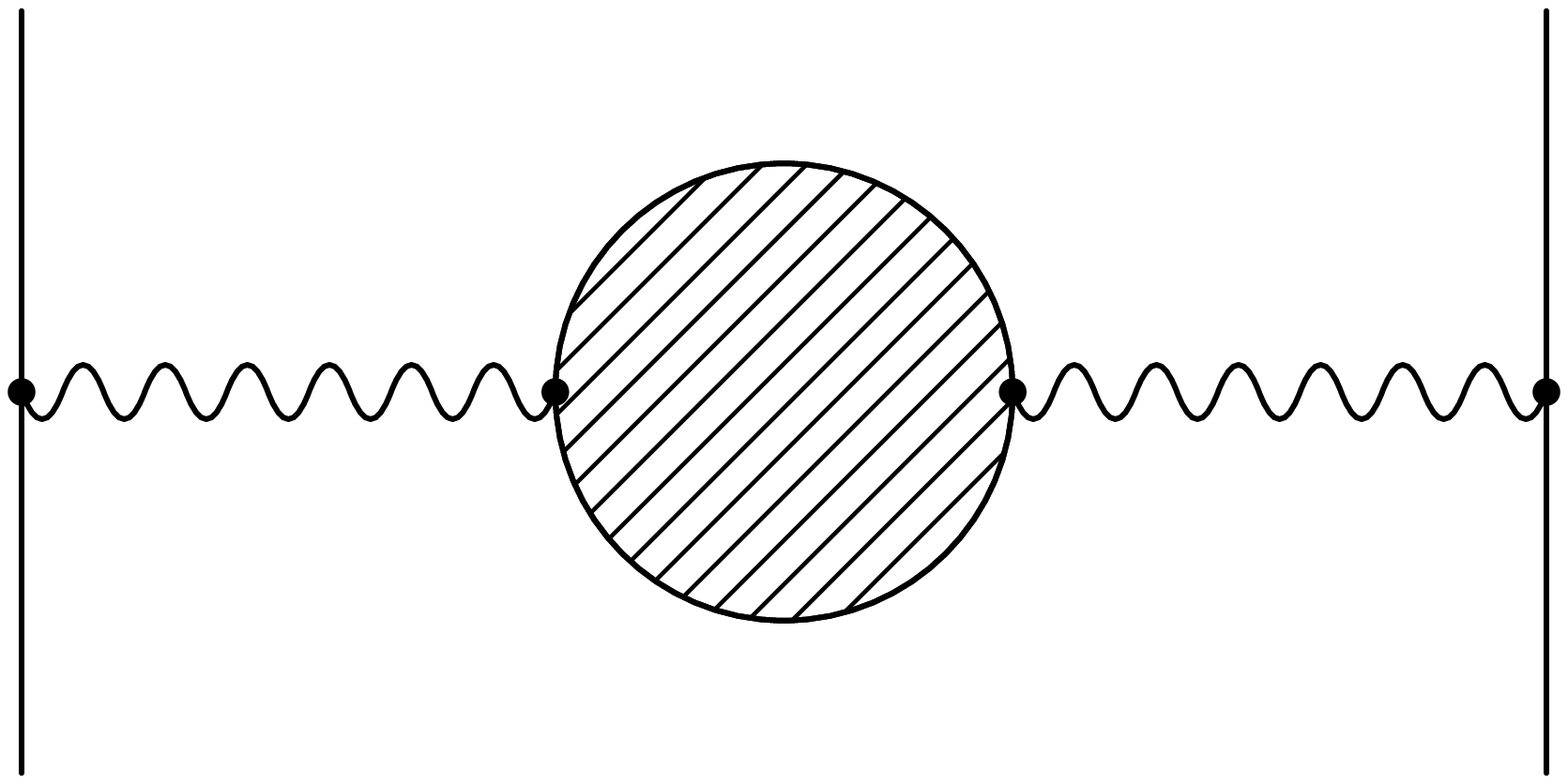}

\vspace{0.2cm}
(a)
\vspace{0.5cm}

\hspace{0.2cm}
\epsfxsize=10cm
\epsffile{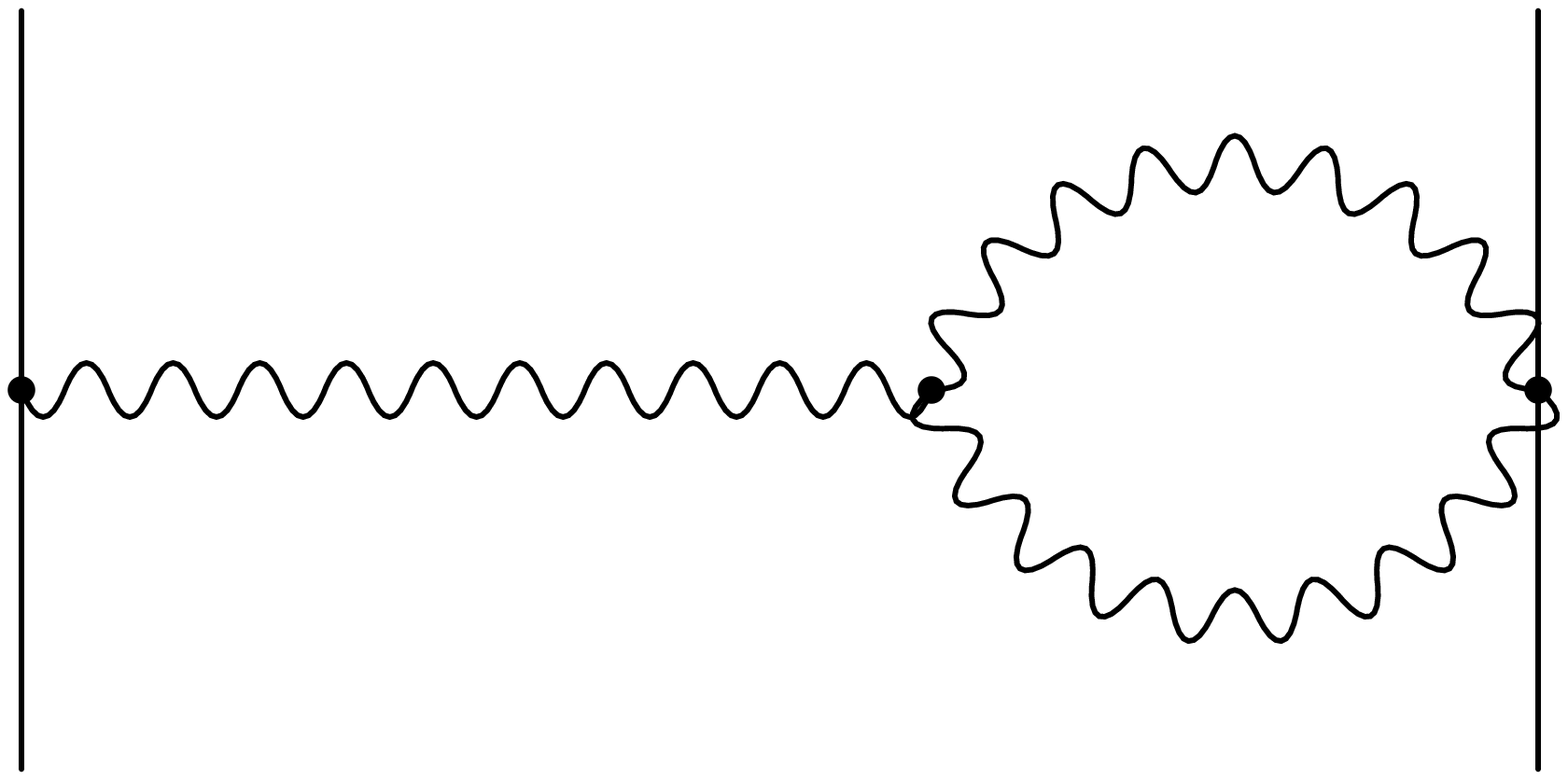}

\vspace{0.2cm}
(b)
\vspace{0.5cm}

\hspace{0.2cm}
\epsfxsize=10cm
\epsffile{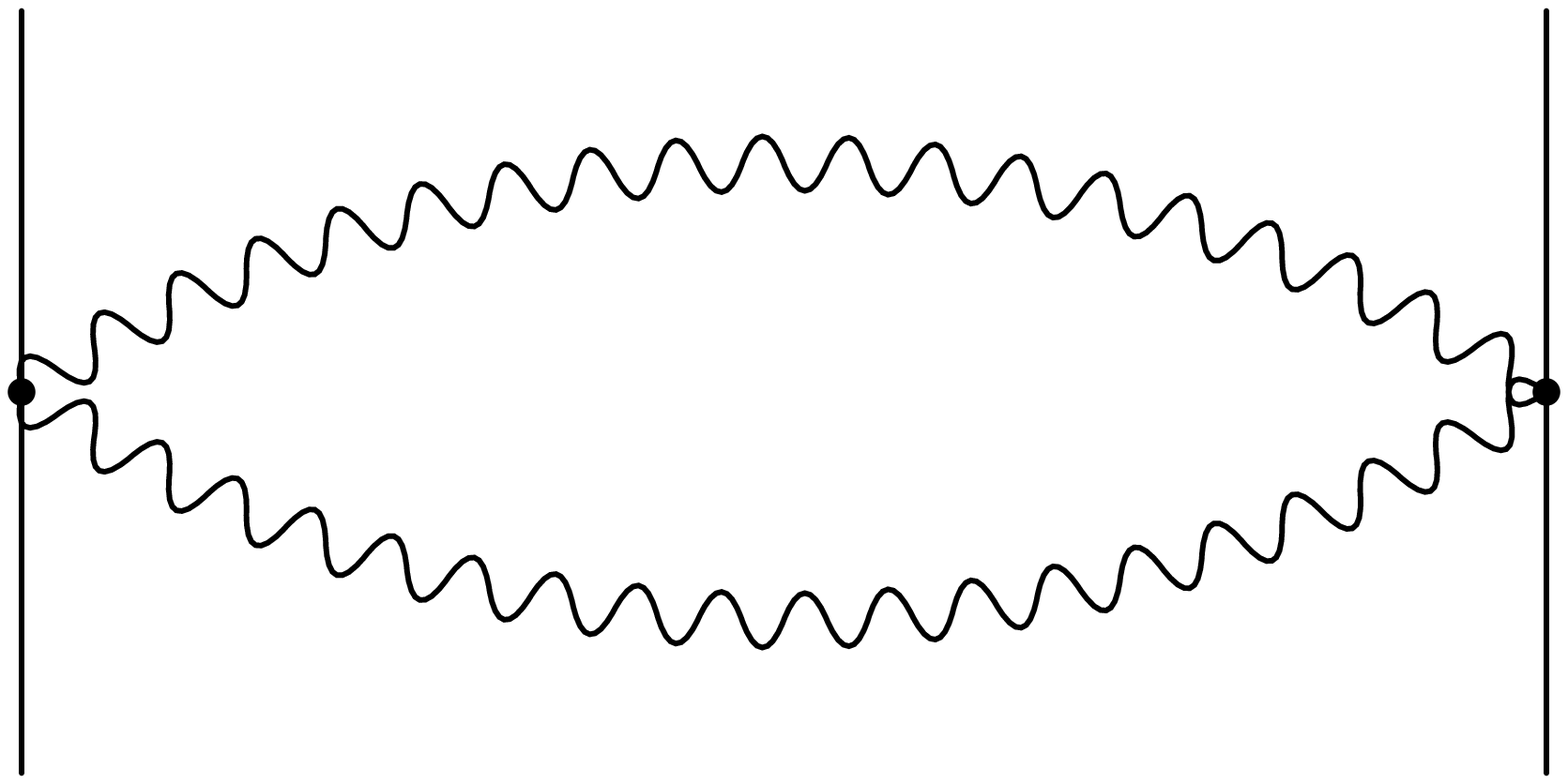}

\vspace{0.2cm}
(c)
\vspace{0.2cm}

\Large{Fig.1}

\end{center}


\newpage
\begin{center}

\hspace{0.2cm}
\epsfxsize=10cm
\epsffile{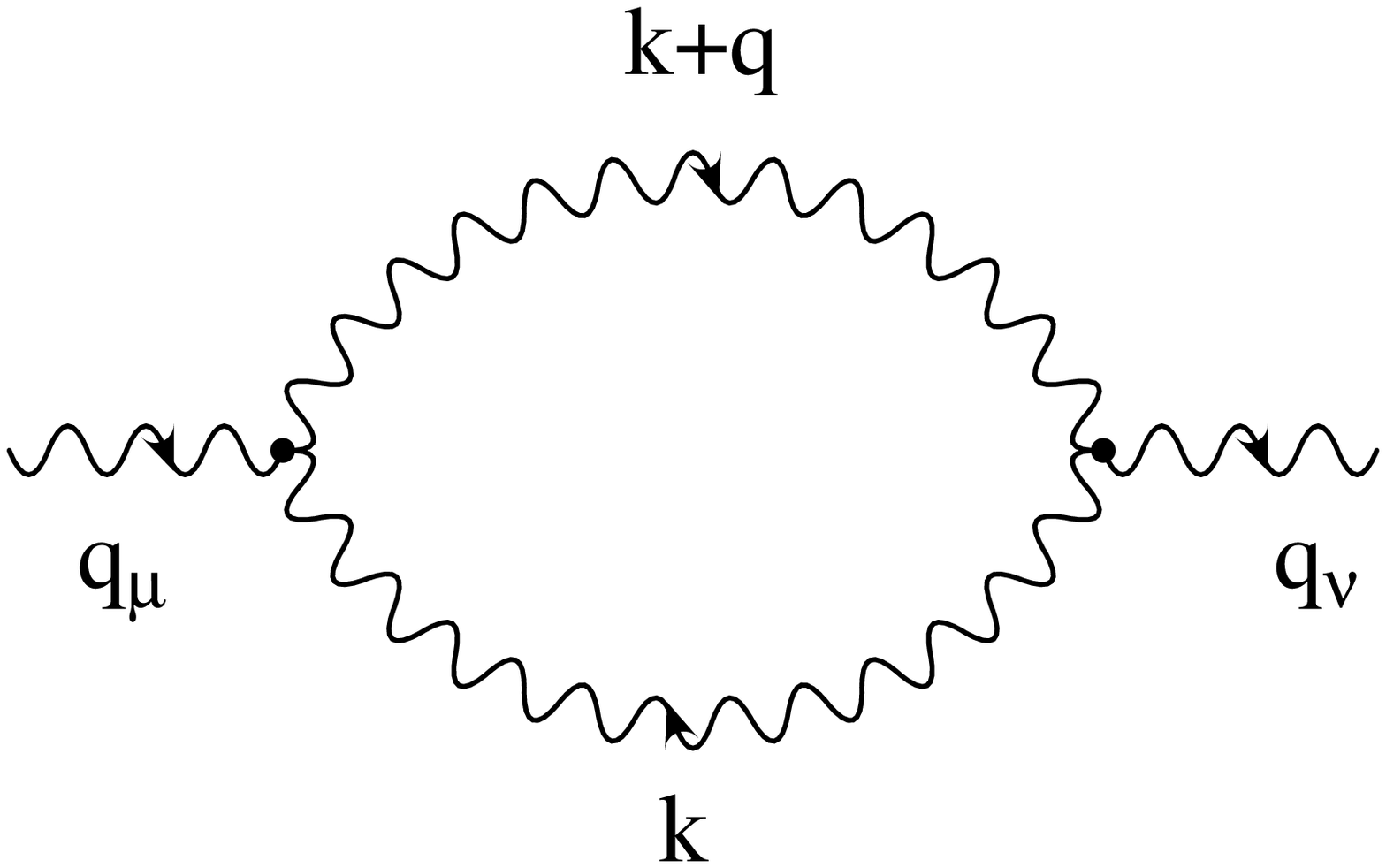}

\vspace{0.2cm}
(a)
\vspace{1cm}

\hspace{0.2cm}
\epsfxsize=10cm
\epsffile{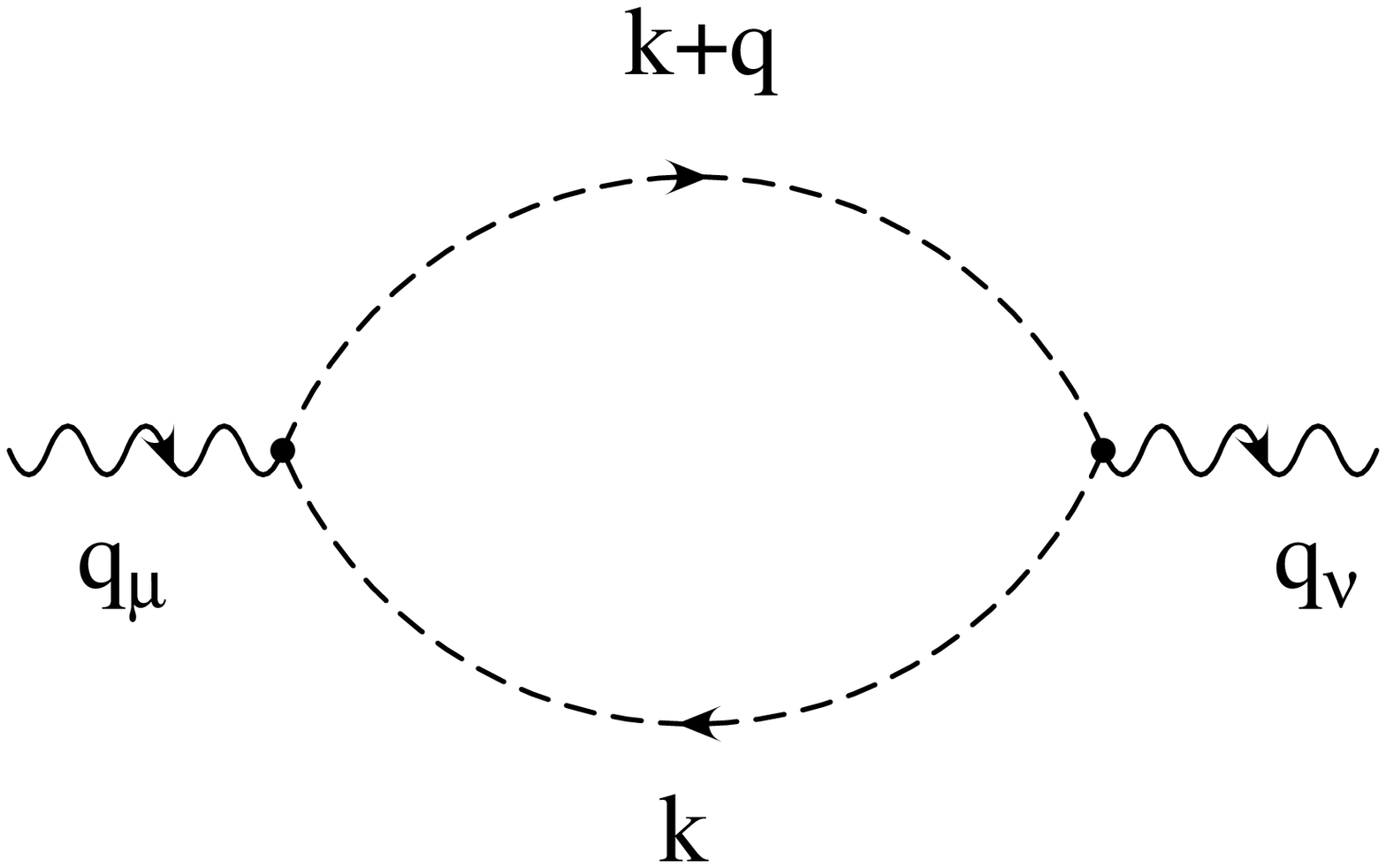}

\vspace{0.2cm}
(b)
\vspace{1cm}

\Large{Fig.2}

\end{center}


\newpage
\vspace*{-1cm}
\begin{center}

\hspace{0.2cm}
\epsfxsize=6.5cm
\epsffile{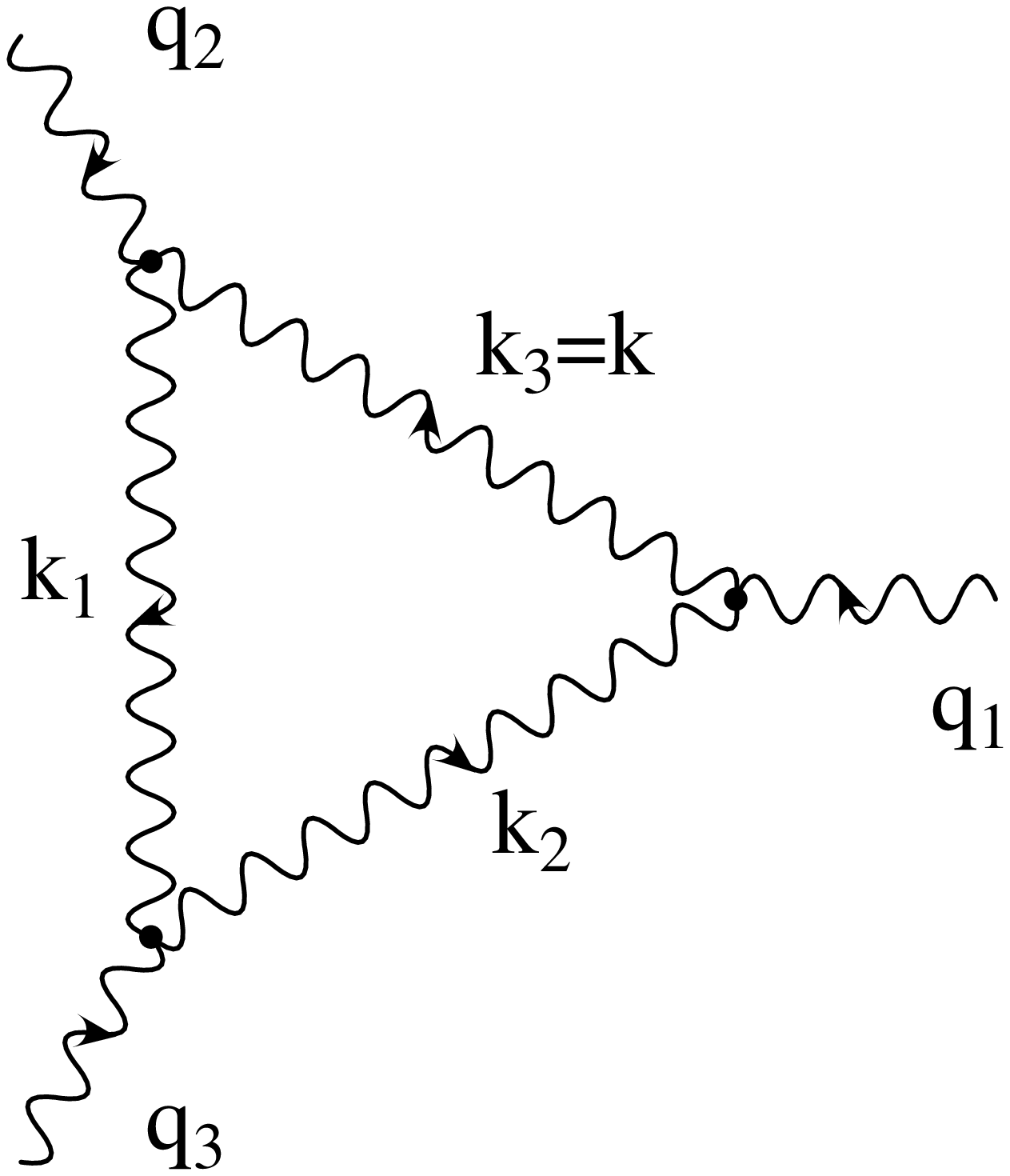}

\vspace{0.2cm}
(a)
\vspace{0.5cm}

\begin{tabular}{cc}
\hspace{0.2cm}
\epsfxsize=6.5cm
\epsffile{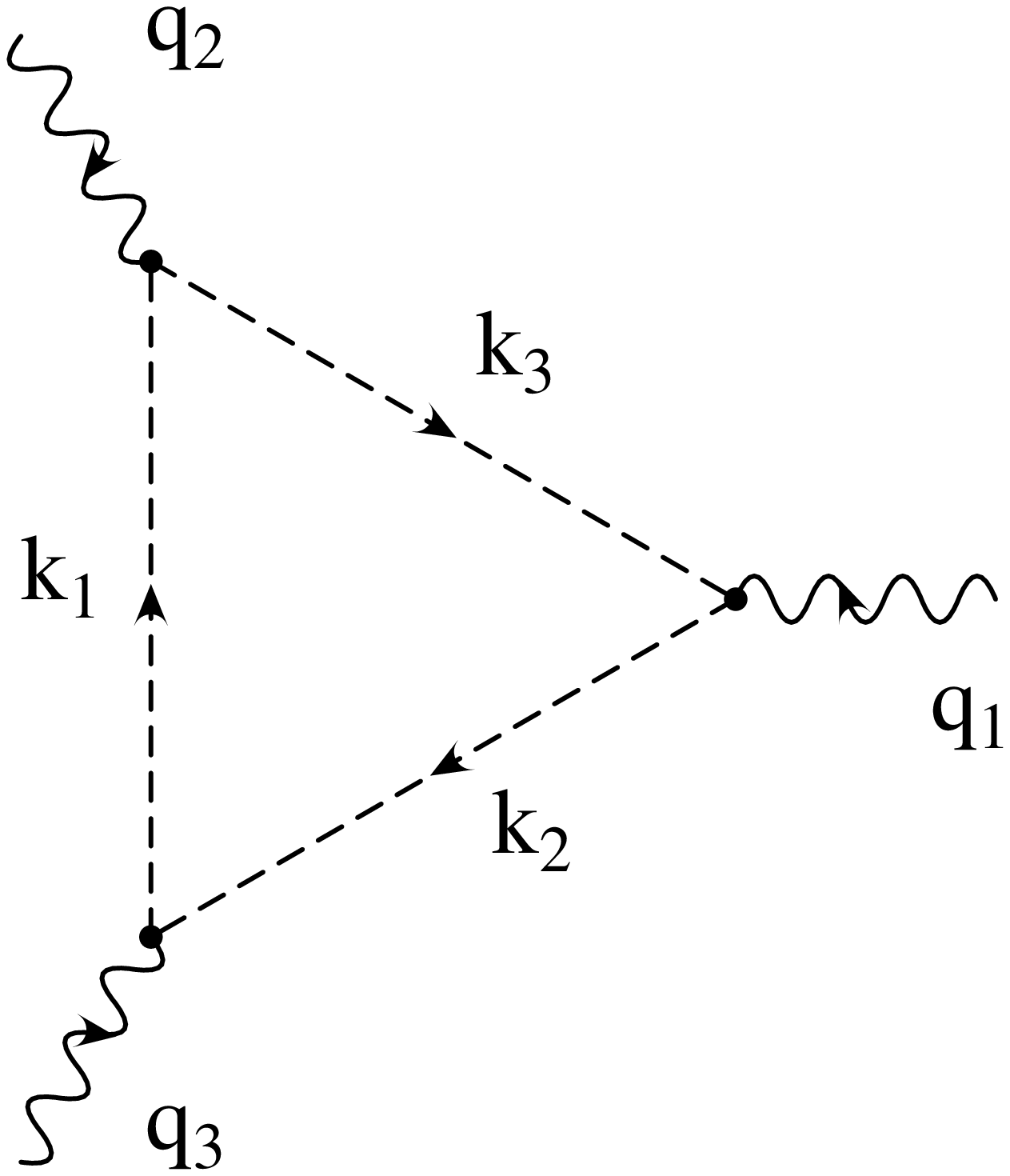}
&
\hspace{0.2cm}
\epsfxsize=6.5cm
\epsffile{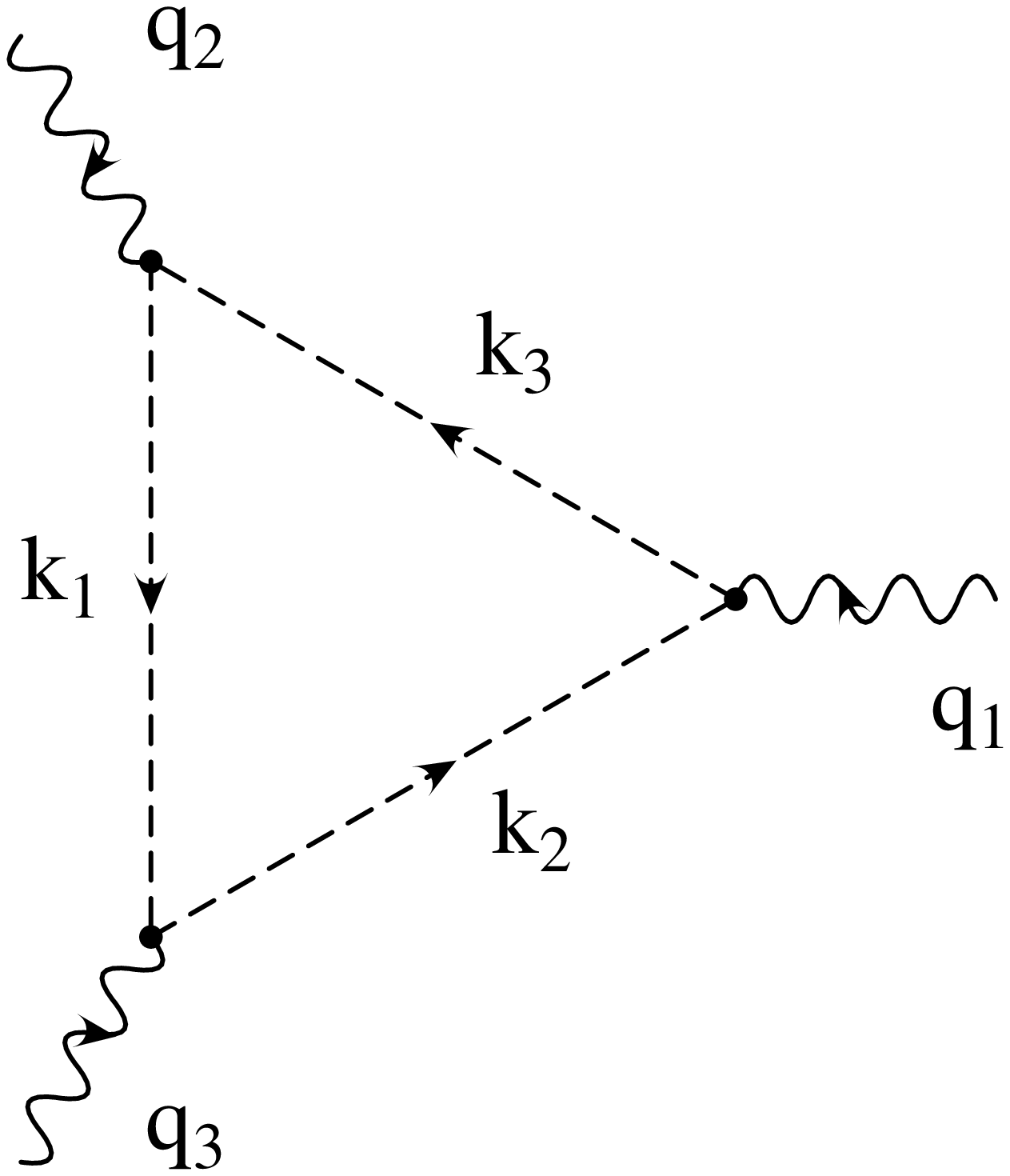}
\\
(b) & (c)
\end{tabular}

\vspace{5mm}
\Large{Fig.3}
\end{center}


\begin{center}
\begin{tabular}{ll}
\epsfxsize=5cm
\epsffile{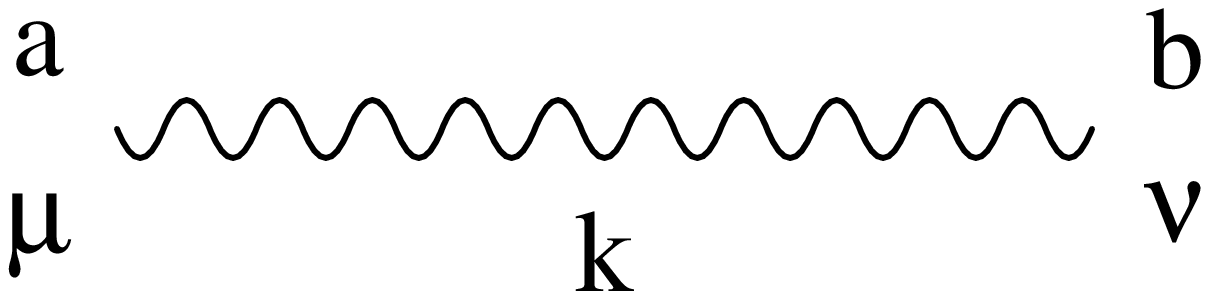}
&
\raisebox{1cm}{
$ \displaystyle
\frac{-i \delta_{ab}}{k^{2} + i \varepsilon}
\left[ g_{\mu \nu} - (1-\xi) \frac{k_{\mu} k_{\nu}}{k^{2}} \right] $}
\\[0.5cm]
\epsfxsize=5cm
\epsffile{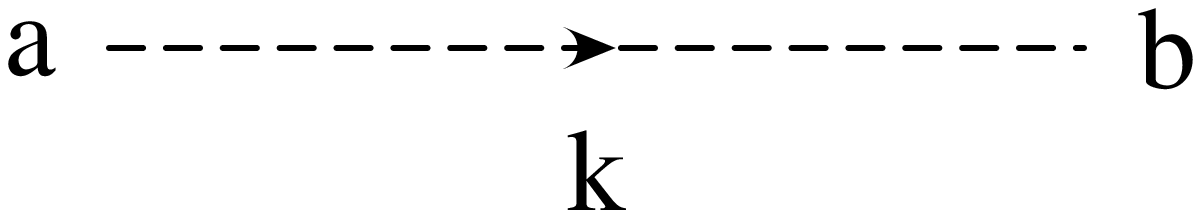}
&
\raisebox{1cm}{
$ \displaystyle
\frac{i \delta_{ab}}{k^{2} + i \varepsilon} $}
\\[0.5cm]
\epsfxsize=5cm
\epsffile{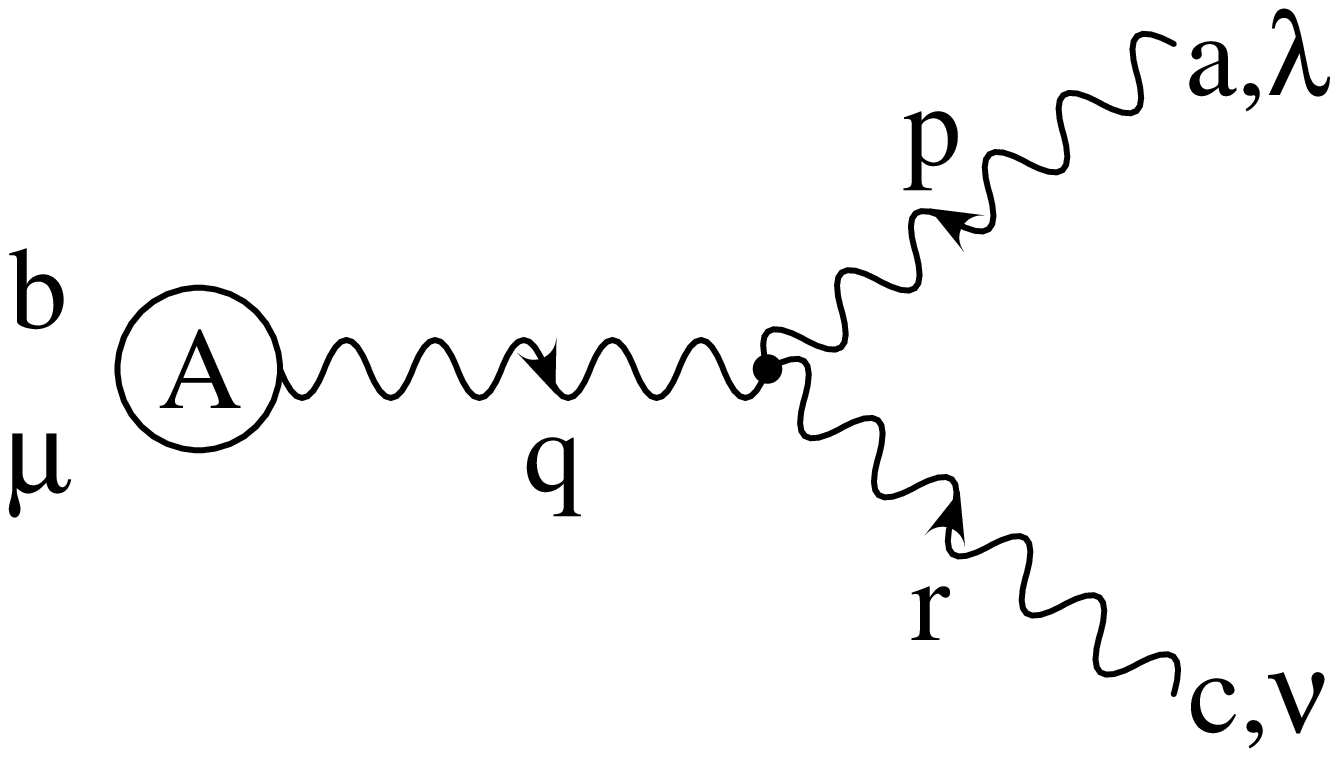}
&
\raisebox{1.5cm}{
$ \displaystyle
g f^{abc} \left[ (p-q+\frac{1}{\xi}r)_{\nu} g_{\lambda \mu}
                +(q-r-\frac{1}{\xi}p)_{\lambda} g_{\mu \nu}
                +(r-p)_{\mu} g_{\nu \lambda} \right] $}
\\[0.5cm]
\epsfxsize=5cm
\epsffile{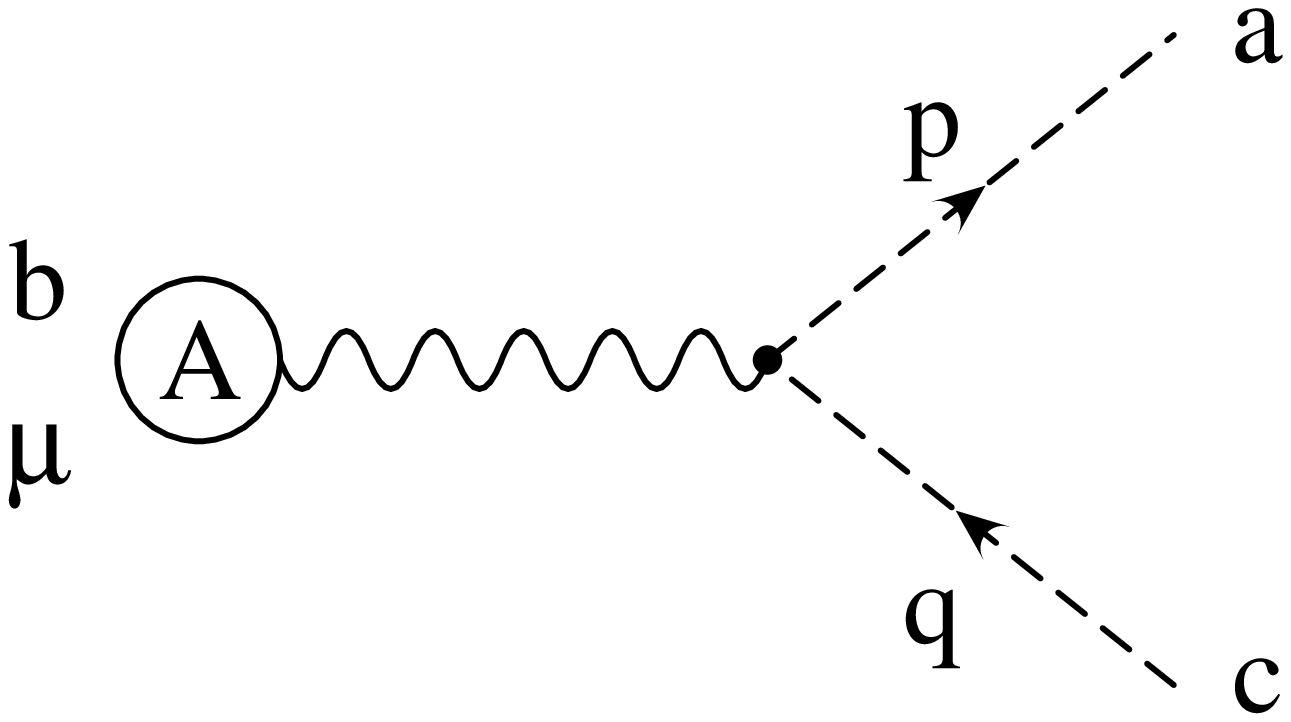}
&
\raisebox{1.5cm}{
$ \displaystyle -g f^{abc} (p+q)_{\mu} $}
\\[0.5cm]
\epsfxsize=5cm
\epsffile{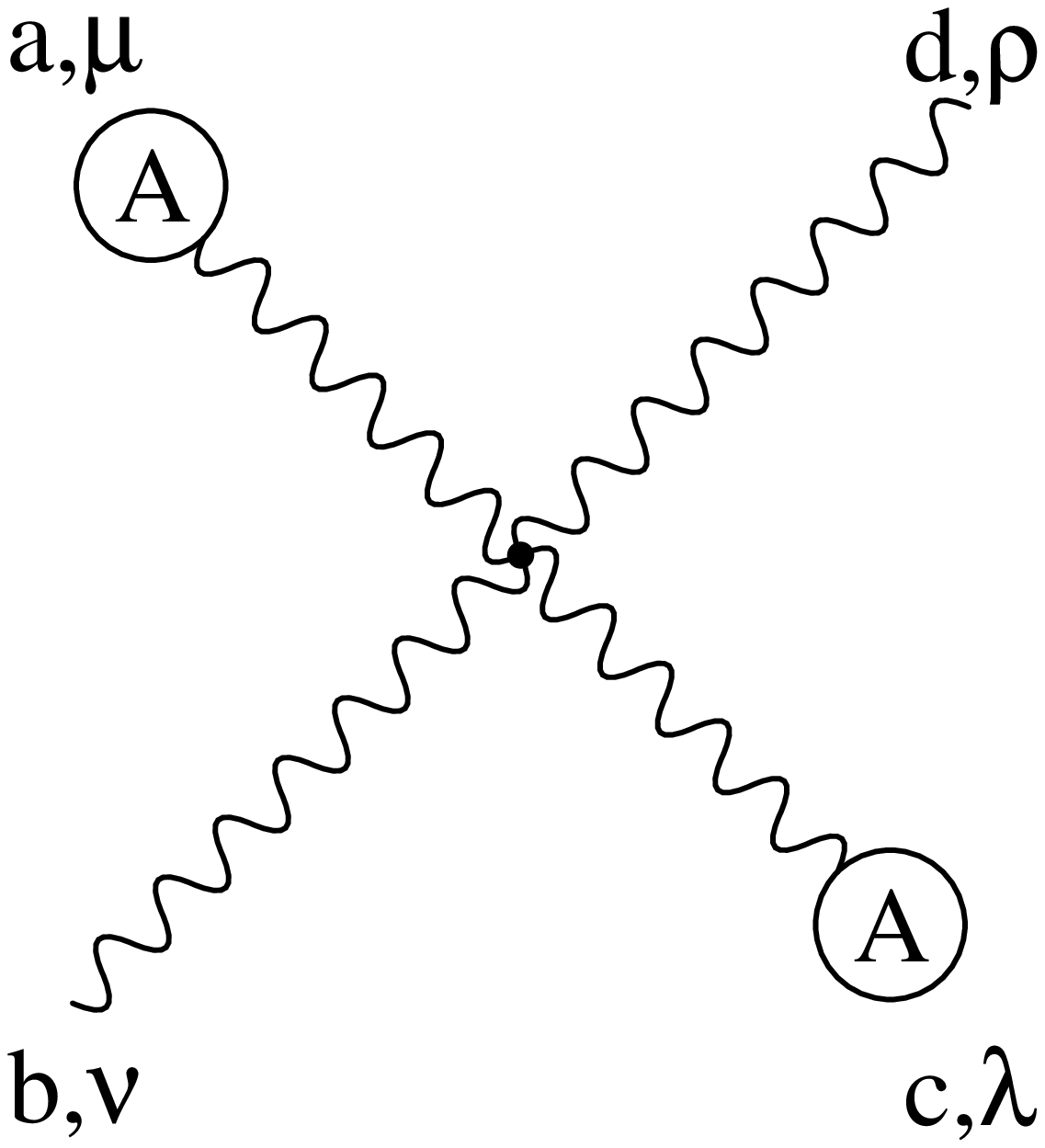}
&
\raisebox{3cm}{
$ 
\begin{array}{ll}
-ig^{2} & [ f^{abx}f^{xcd} (  g_{\mu \lambda} g_{\nu \rho}
                           - g_{\mu \rho} g_{\nu \lambda}
                           + \frac{1}{\xi} g_{\mu \nu} g_{\lambda \rho} ) \\
        &  +f^{adx}f^{xbc} (  g_{\mu \nu} g_{\lambda \rho}
                           - g_{\mu \lambda} g_{\nu \rho}
                           - \frac{1}{\xi} g_{\mu \rho} g_{\nu \lambda} ) \\
        &  +f^{acx}f^{xbd} (  g_{\mu \nu} g_{\lambda \rho}
                           - g_{\mu \rho} g_{\nu \lambda} ) ]
\end{array} $ }
\\[0.5cm]
\epsfxsize=5cm
\epsffile{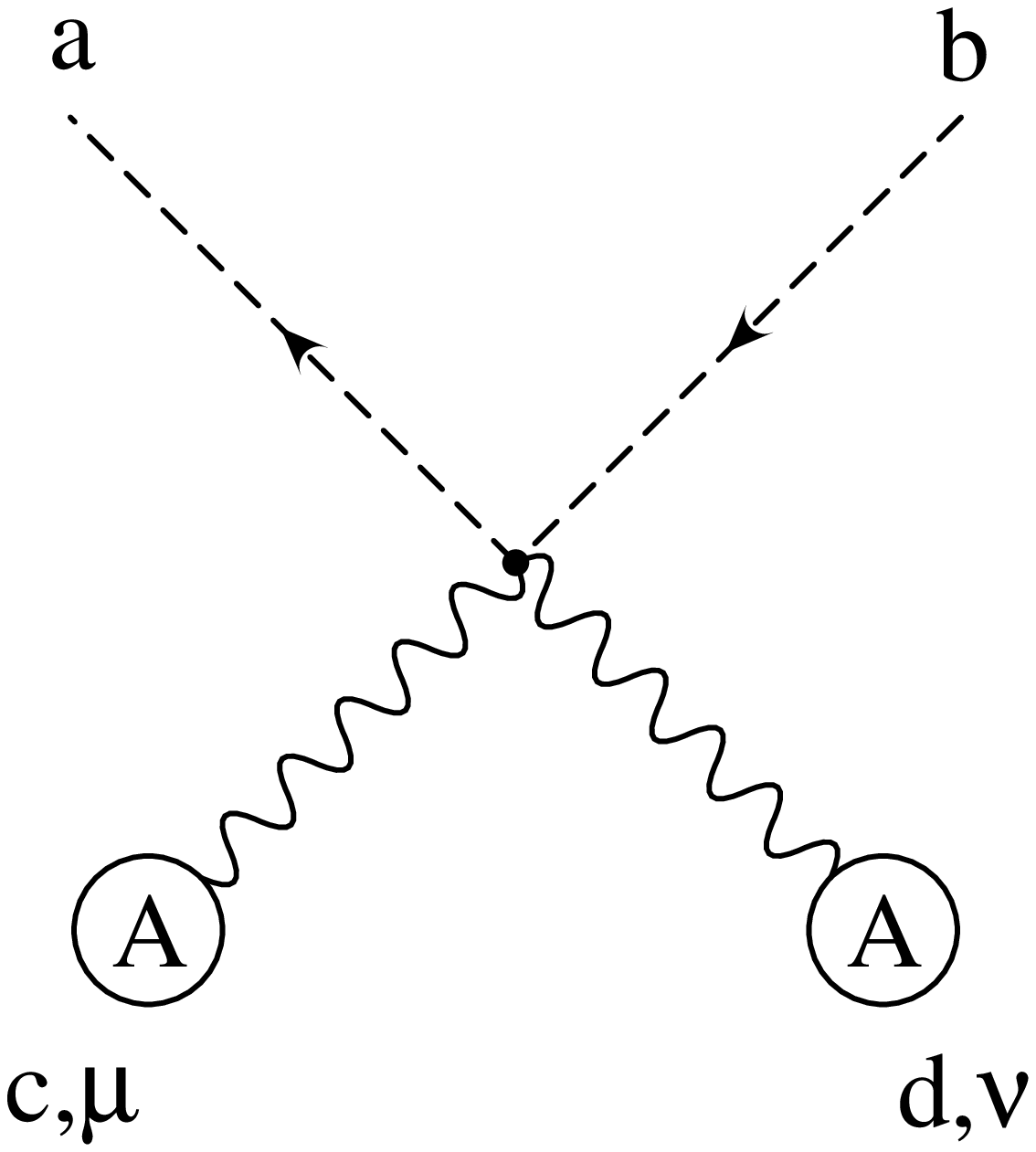}
&
\raisebox{3cm}{
$ -ig^{2} g_{\mu \nu} (f^{acx}f^{xdb} + f^{adx}f^{xcb}) $ }
\end{tabular}
\end{center}


\newpage
\begin{center}

\hspace{0.2cm}
\epsfxsize=10cm
\epsffile{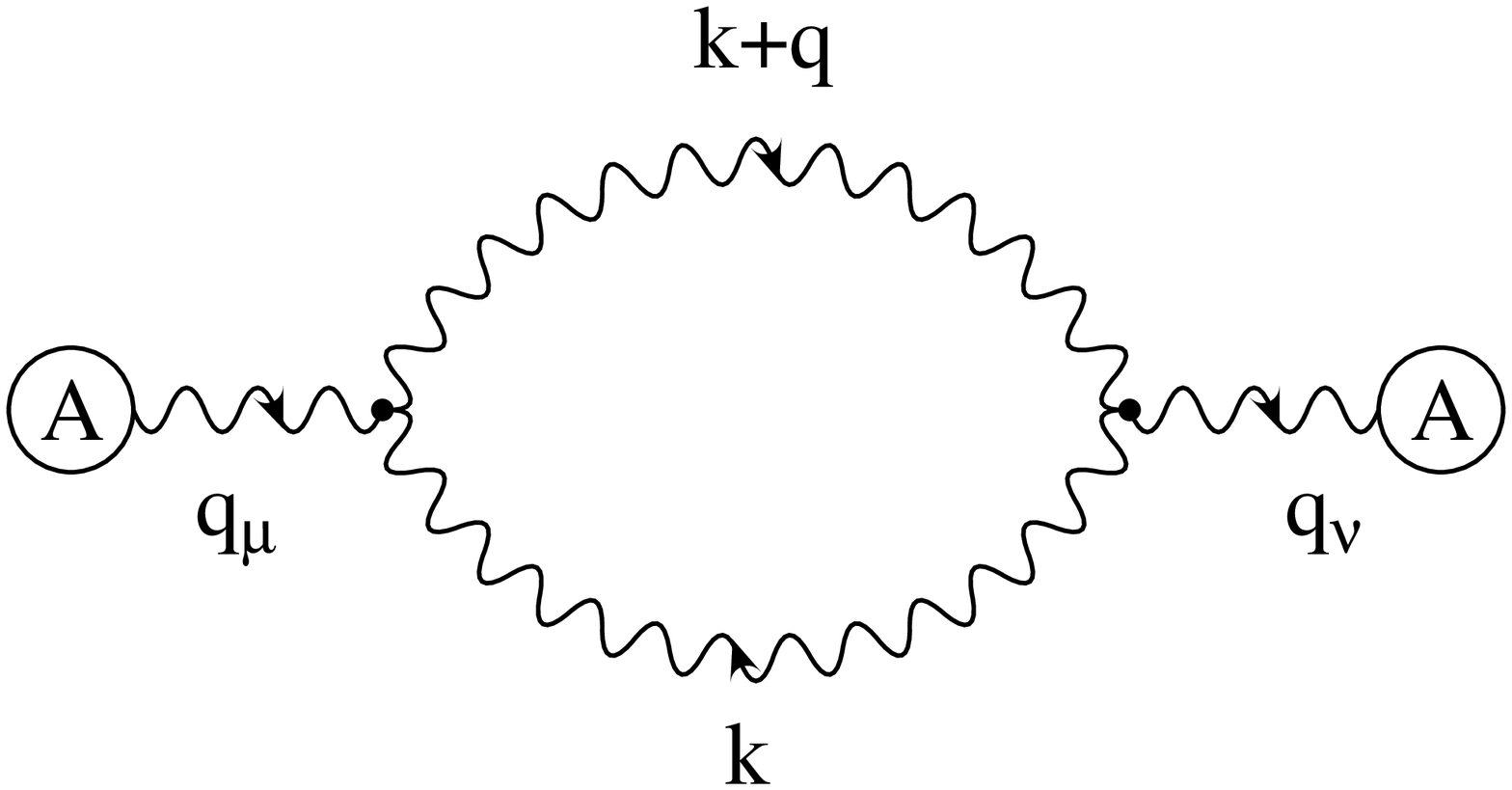}

\vspace{0.2cm}
(a)
\vspace{1cm}

\hspace{0.2cm}
\epsfxsize=10cm
\epsffile{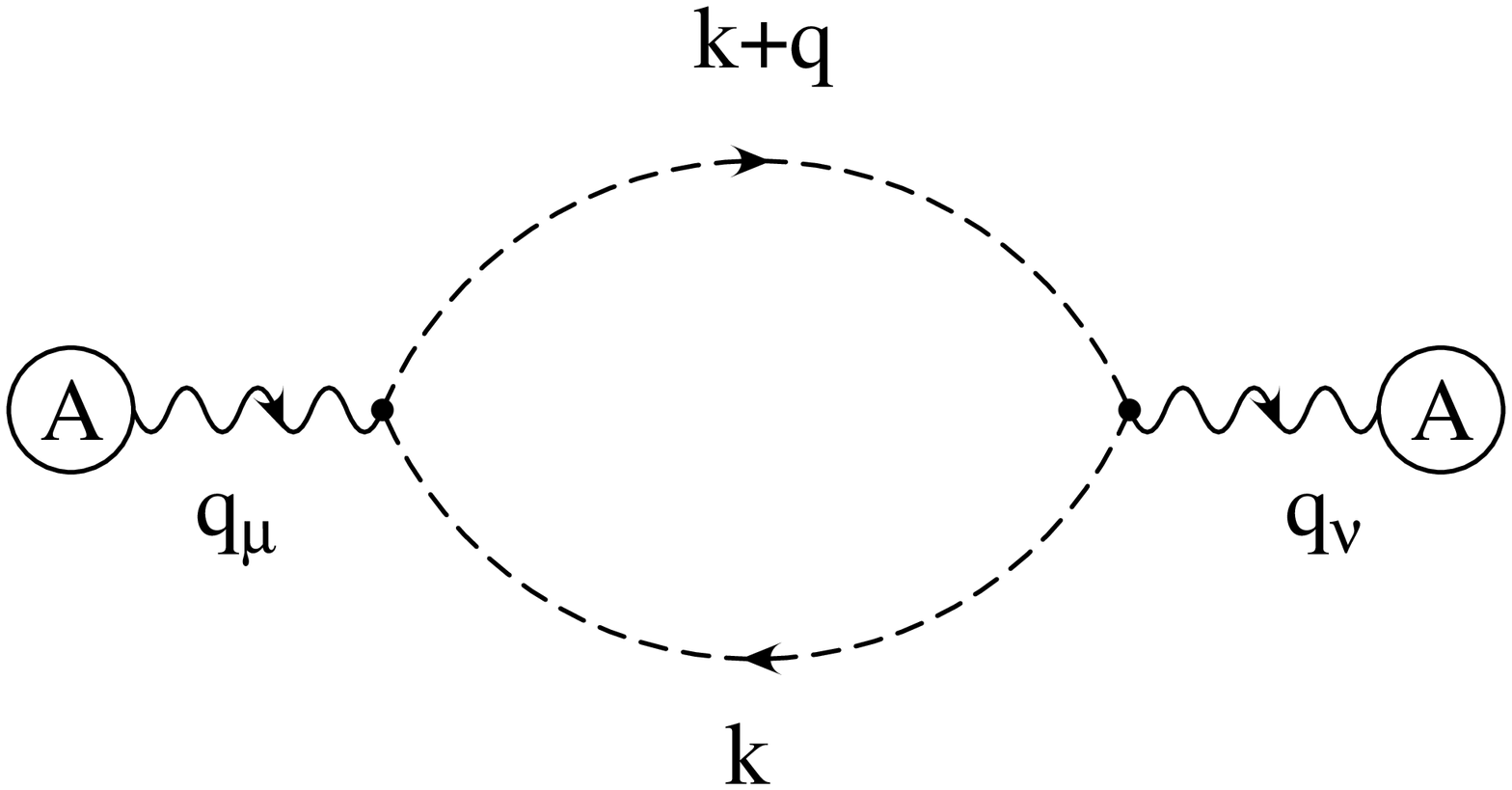}

\vspace{0.2cm}
(b)
\vspace{1cm}

\Large{Fig.5}

\end{center}


\newpage
\begin{center}

\hspace{0.2cm}
\epsfxsize=6cm
\epsffile{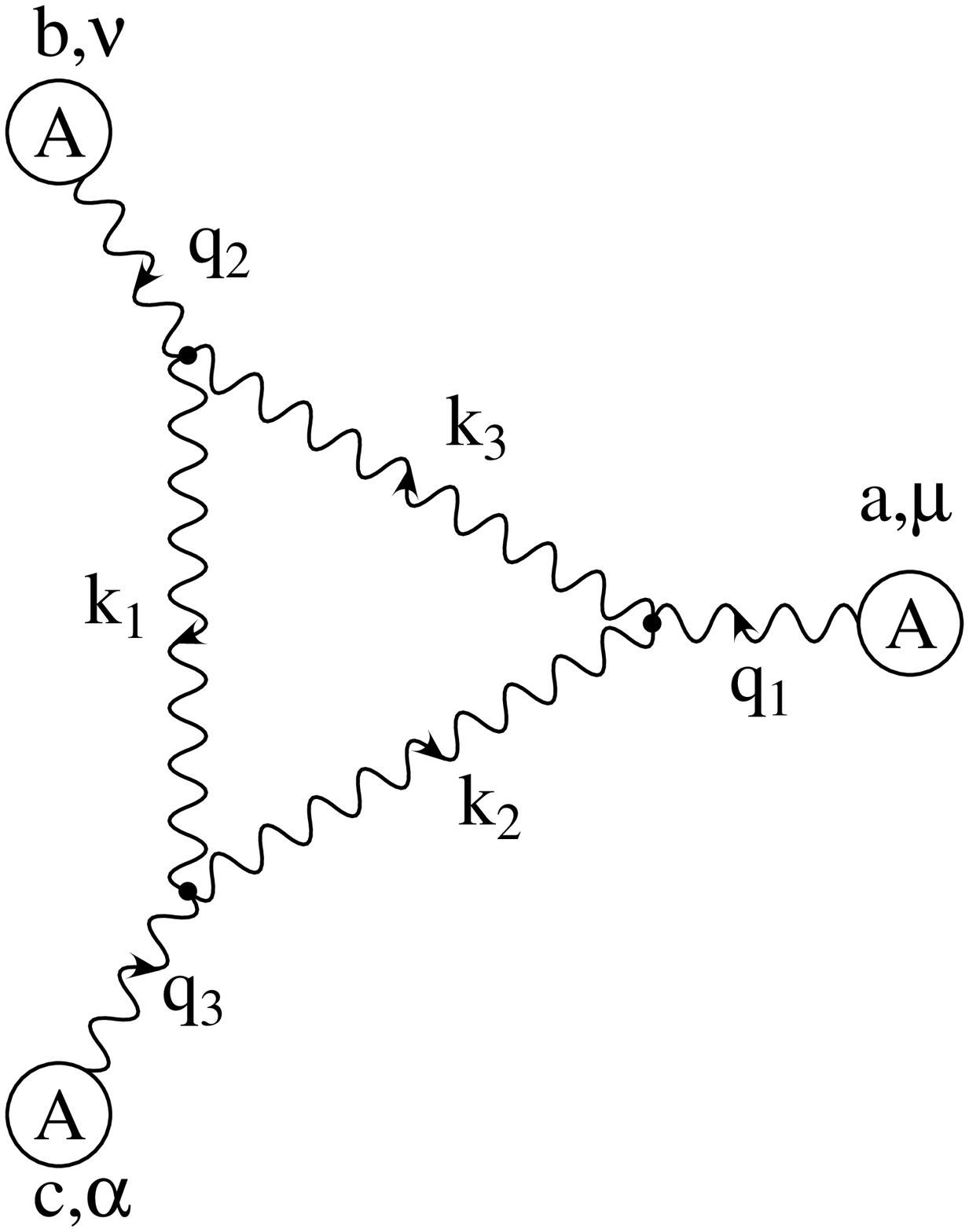}

\vspace{0.2cm}
(a)
\vspace{1cm}

\hspace{0.2cm}
\epsfxsize=6cm
\epsffile{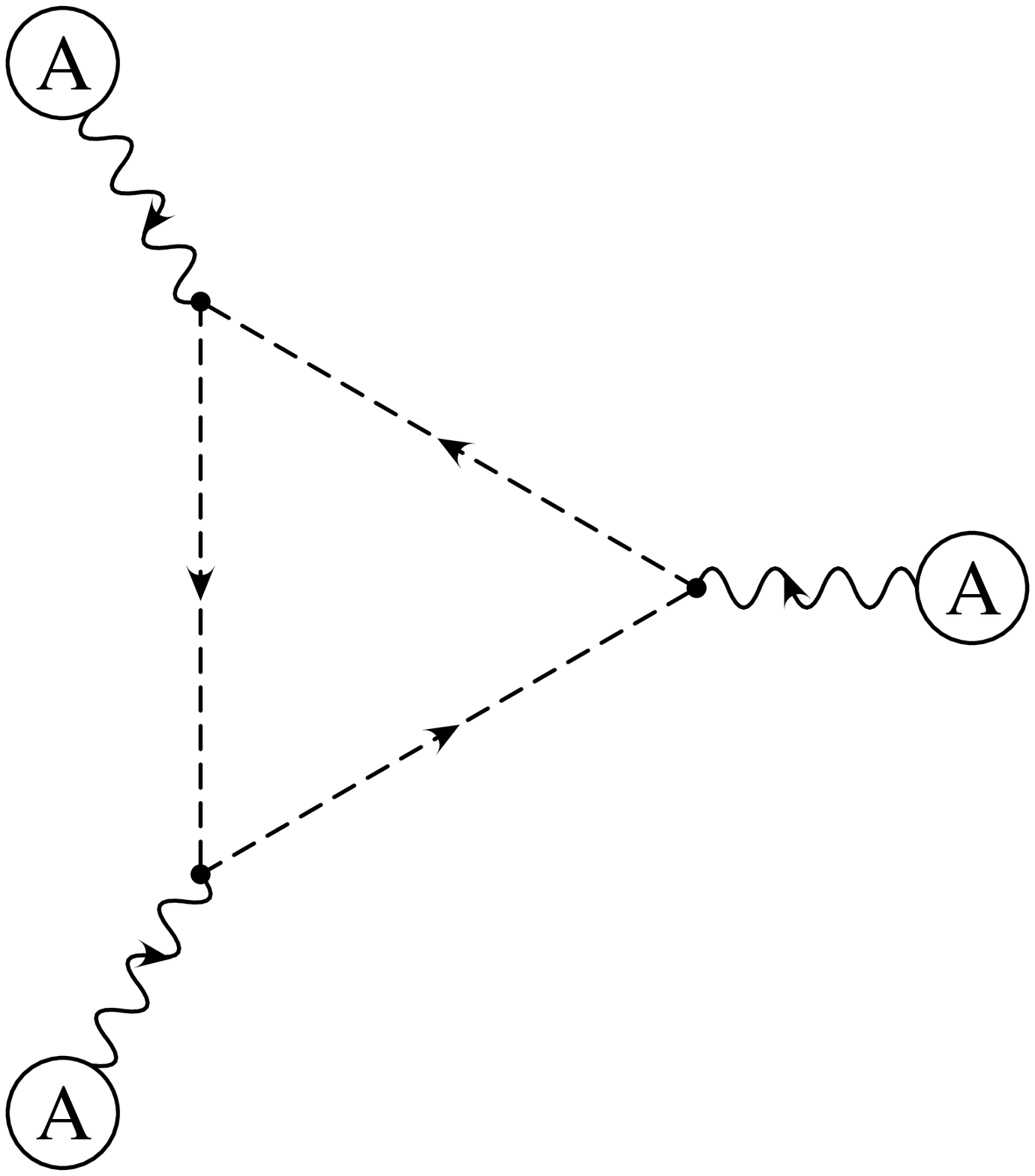}

\vspace{0.2cm}
(b)

\newpage

\begin{tabular}{ll}
\hspace{0.2cm}
\epsfxsize=6cm
\epsffile{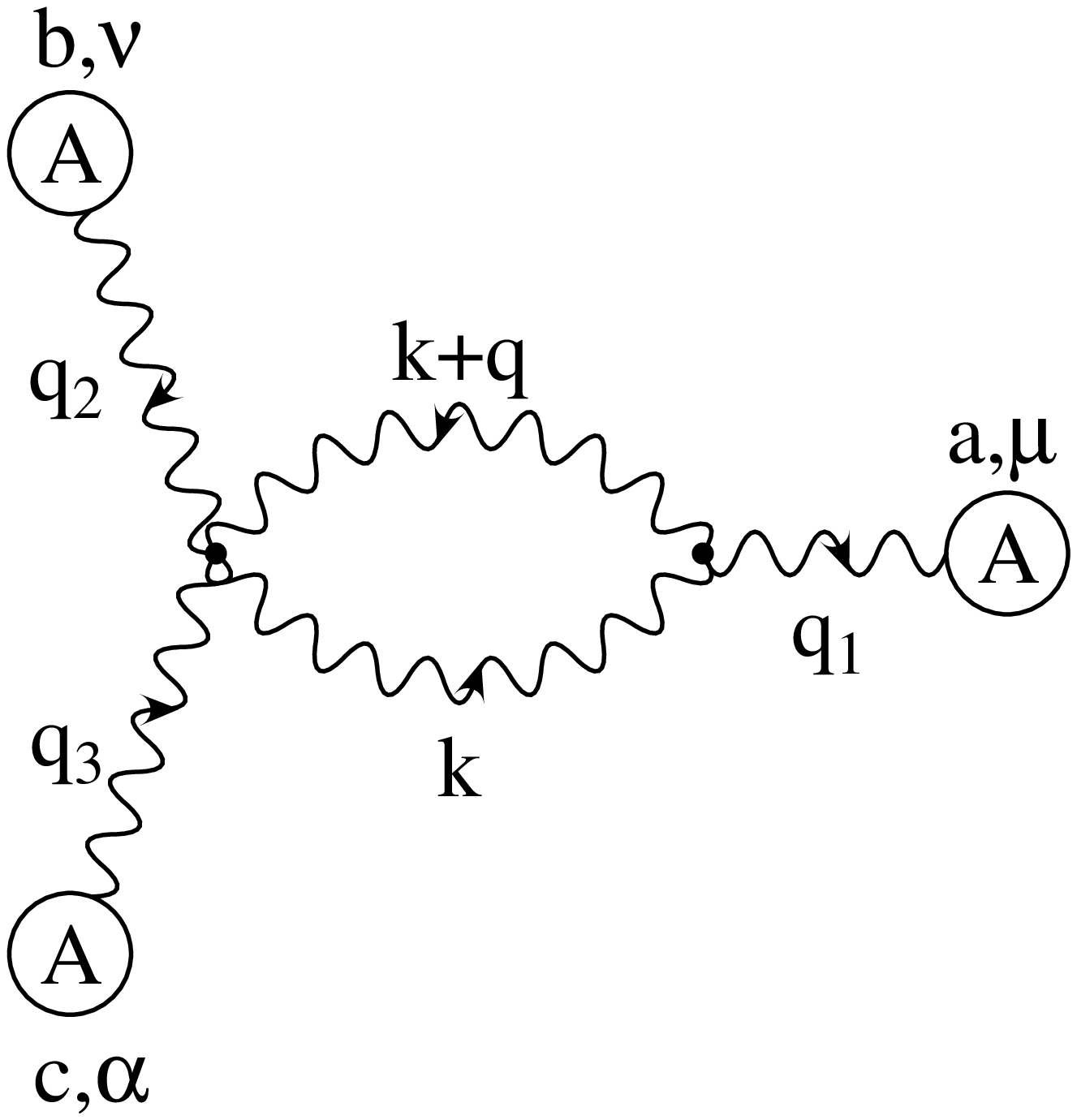}
&
\raisebox{3.2cm}{+ 2 permutations} \\[0.2cm]
\multicolumn{2}{c}{(c)} \\[1cm]

\hspace{0.2cm}
\epsfxsize=6cm
\epsffile{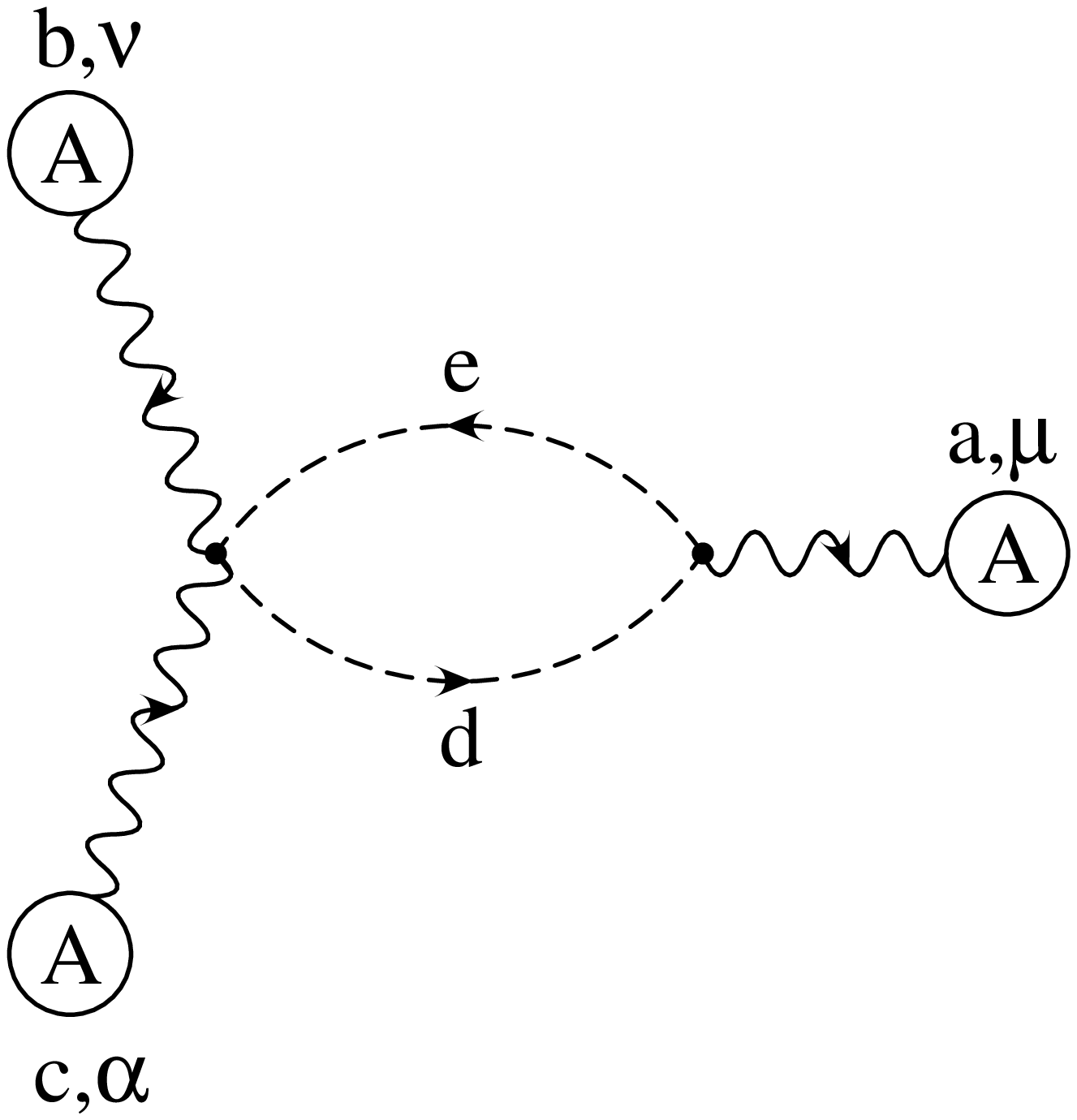}
&
\raisebox{3.2cm}{+ 2 permutations} \\[0.2cm]
\multicolumn{2}{c}{(d)} \\[1cm]
\end{tabular}

\vspace{1cm}
\Large{Fig.6}
\end{center}


\newpage

\vspace*{-1cm}
\begin{center}
\begin{tabular}{ll}
\hspace{0.2cm}
\epsfxsize=7cm
\epsffile{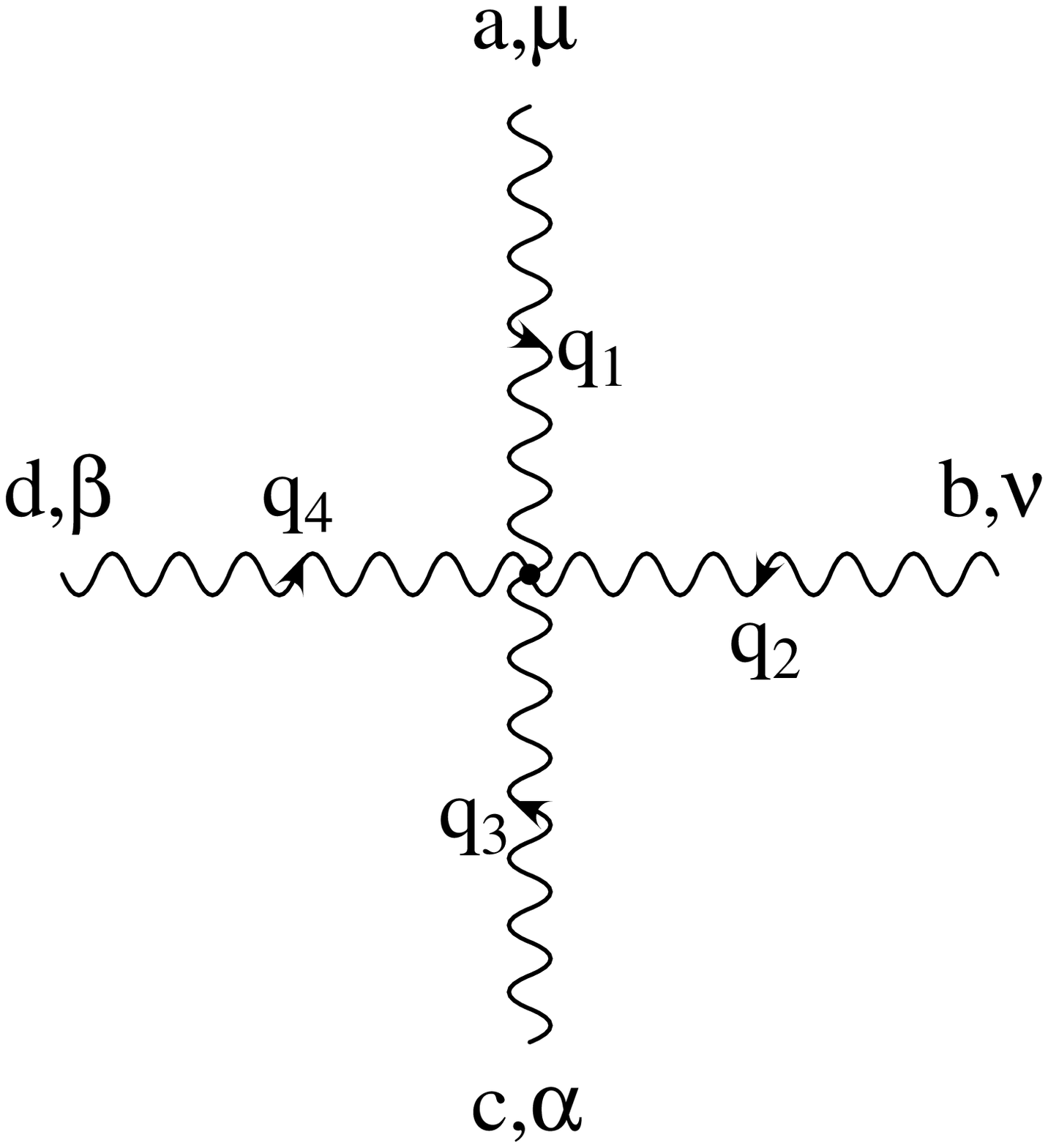}
&
\raisebox{3.2cm}{$ -ig^{2} \Gamma^{abcd}_{\mu \nu \alpha \beta} $} \\[0.2cm]
\multicolumn{2}{c}{(a)} \\[0.5cm]

\hspace{0.2cm}
\epsfxsize=6cm
\epsffile{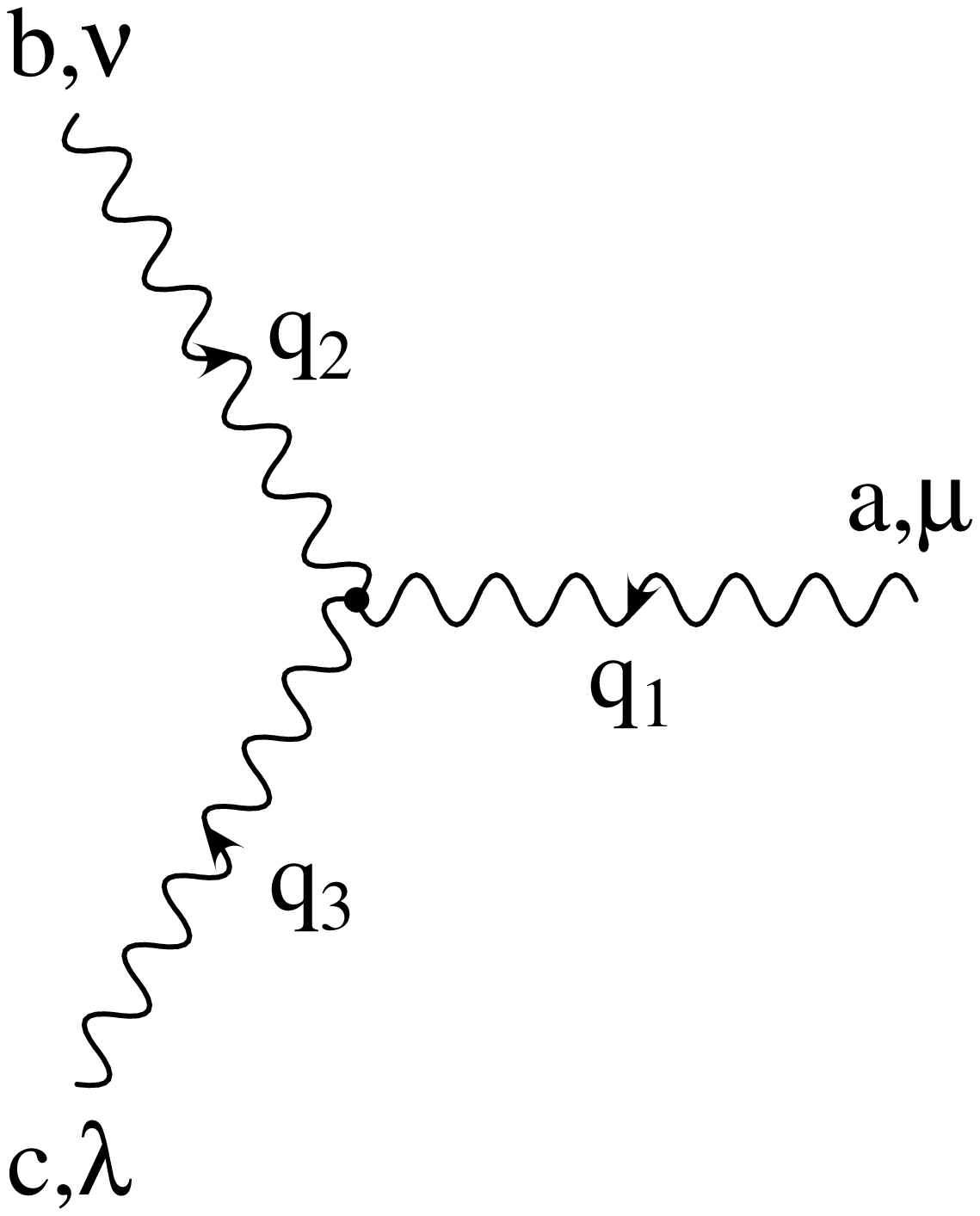}
&
\raisebox{3.2cm}{$ g \Gamma^{abc}_{\mu \nu \lambda}$ } \\[0.2cm]
\multicolumn{2}{c}{(b)} \\[0.5cm]
\end{tabular}

\Large{Fig.7}
\end{center}


\vspace*{-1cm}
\begin{center}
\begin{tabular}{ll}
\hspace{0.2cm}
\epsfxsize=7cm
\epsffile{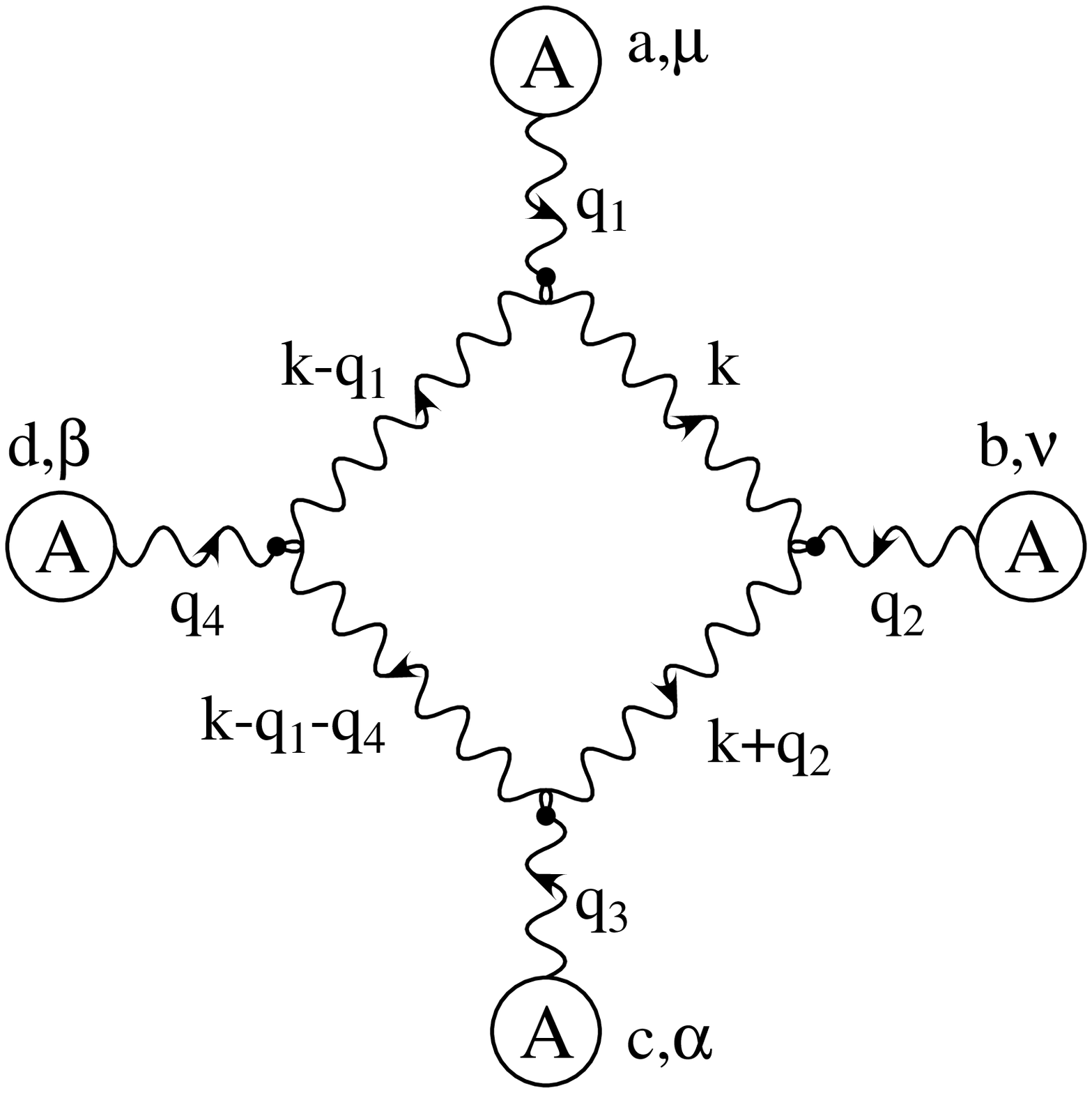}
&
\raisebox{3.2cm}{+ 2 permutations} \\[0.2cm]
\multicolumn{2}{c}{(a)} \\[1cm]

\hspace{0.2cm}
\epsfxsize=6cm
\epsffile{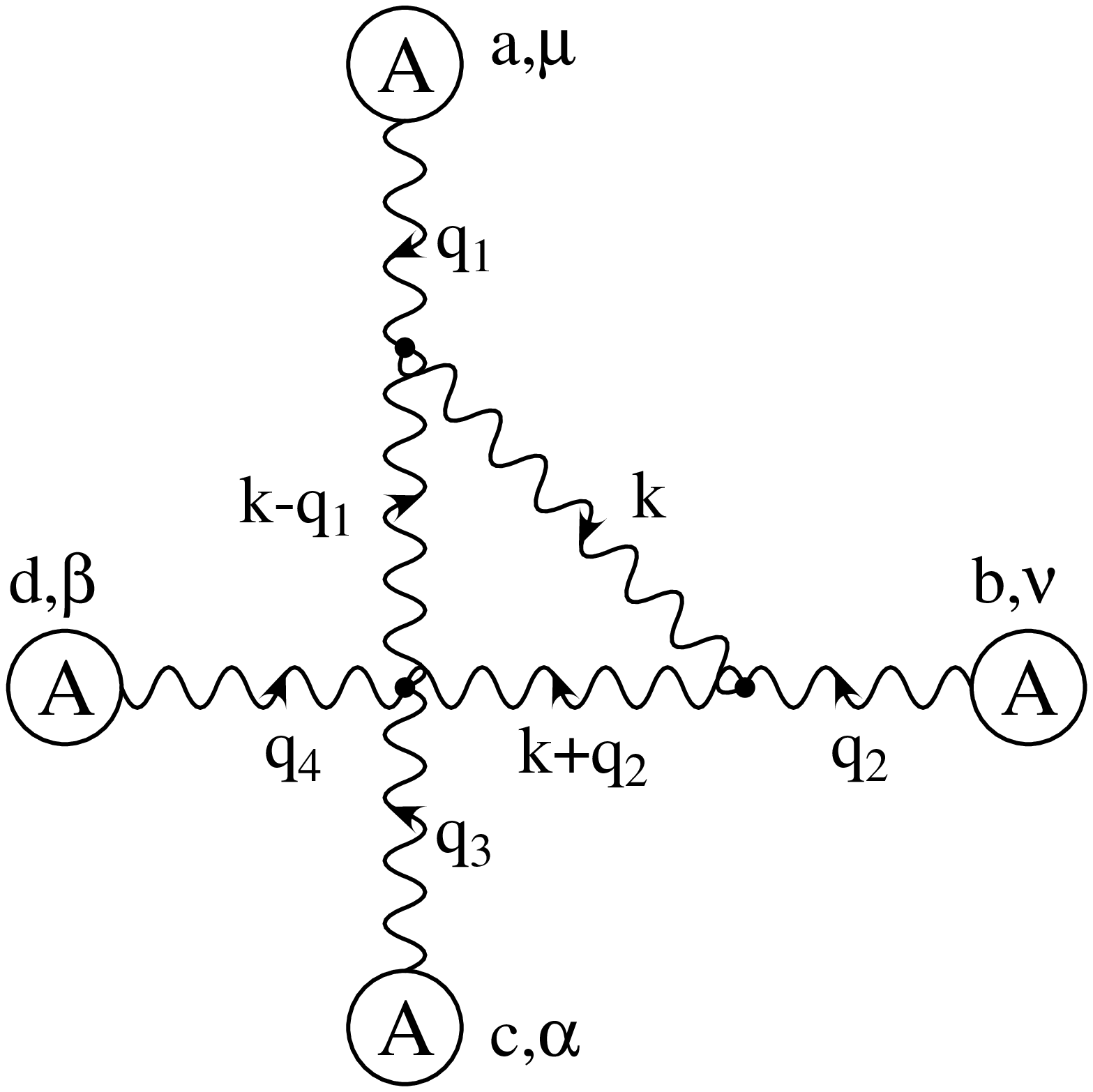}
&
\raisebox{3.2cm}{+ 5 permutations} \\[0.2cm]
\multicolumn{2}{c}{(c)} \\[0.5cm]
\end{tabular}
\end{center}

\newpage
\begin{center}
\begin{tabular}{lll}
\hspace{0.2cm}
&
\epsfxsize=7cm
\epsffile{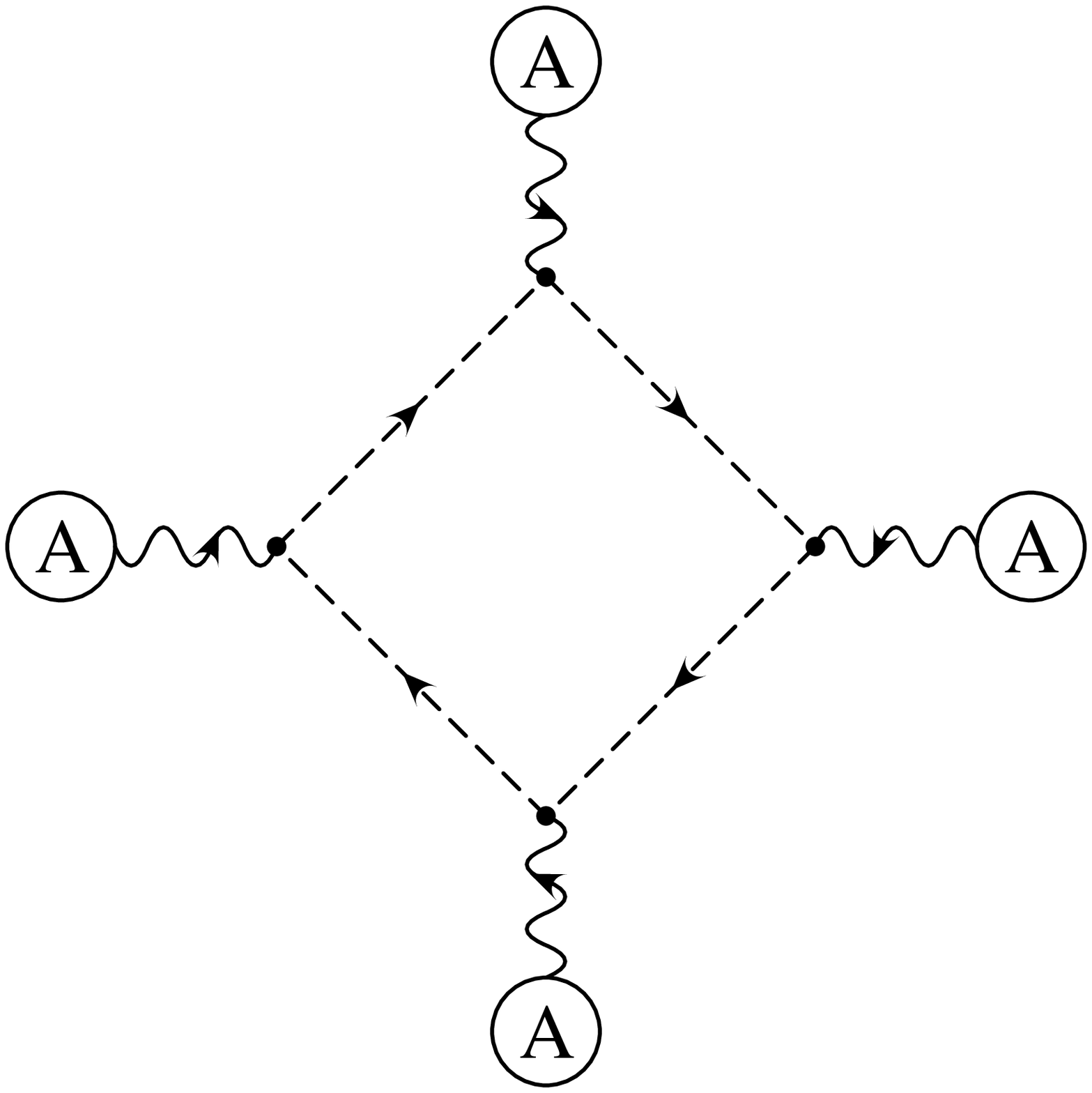}
&
\raisebox{3.6cm}{+ 2 permutations} \\[1cm]
\raisebox{3.6cm}{+} &
\epsfxsize=7cm
\epsffile{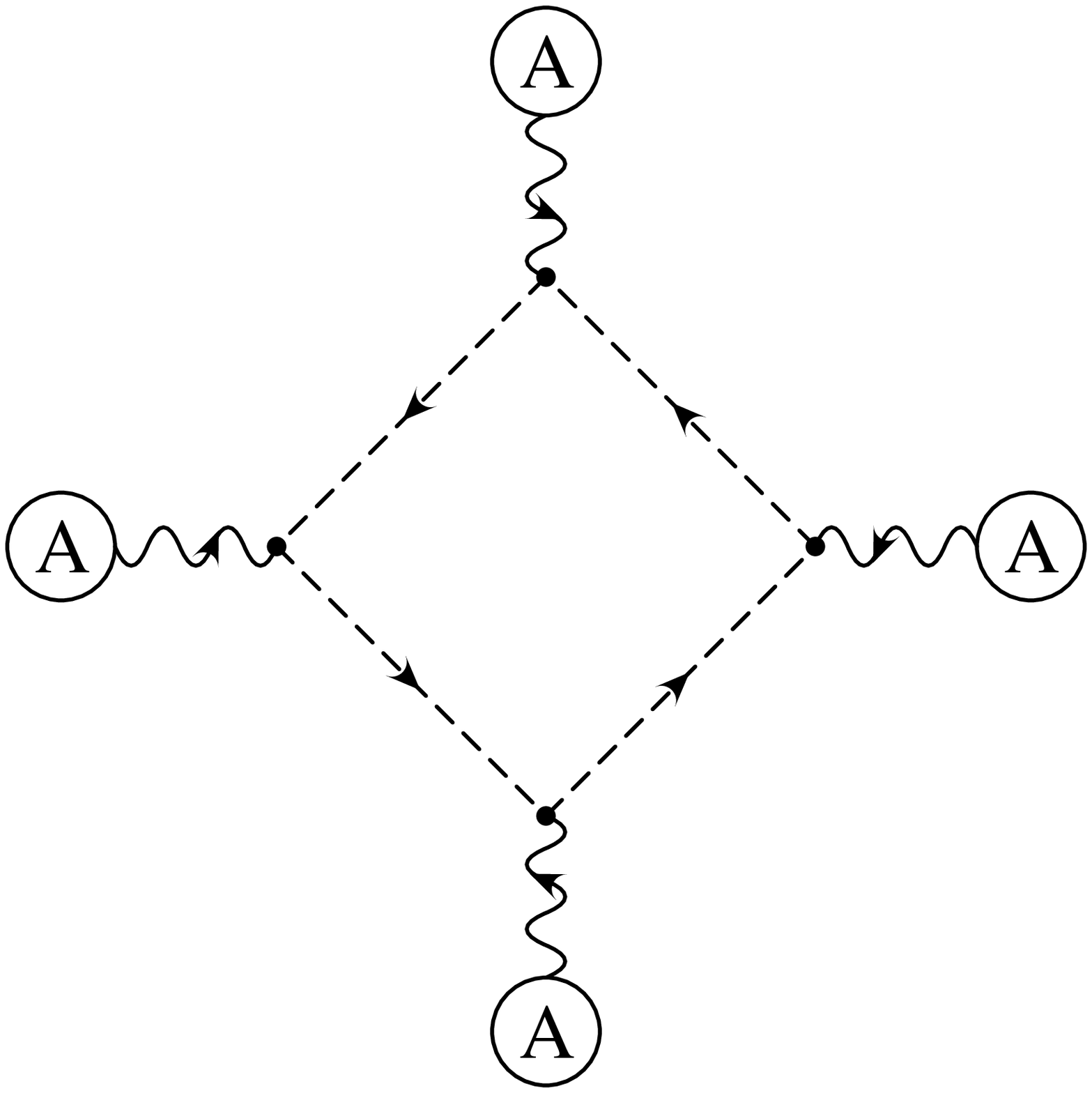}
&
\raisebox{3.6cm}{+ 2 permutations} \\[0.2cm]
\multicolumn{3}{c}{(b)} \\[0.5cm]
\end{tabular}
\end{center}

\newpage
\begin{center}
\begin{tabular}{lll}
\hspace{0.2cm}
&
\epsfxsize=7cm
\epsffile{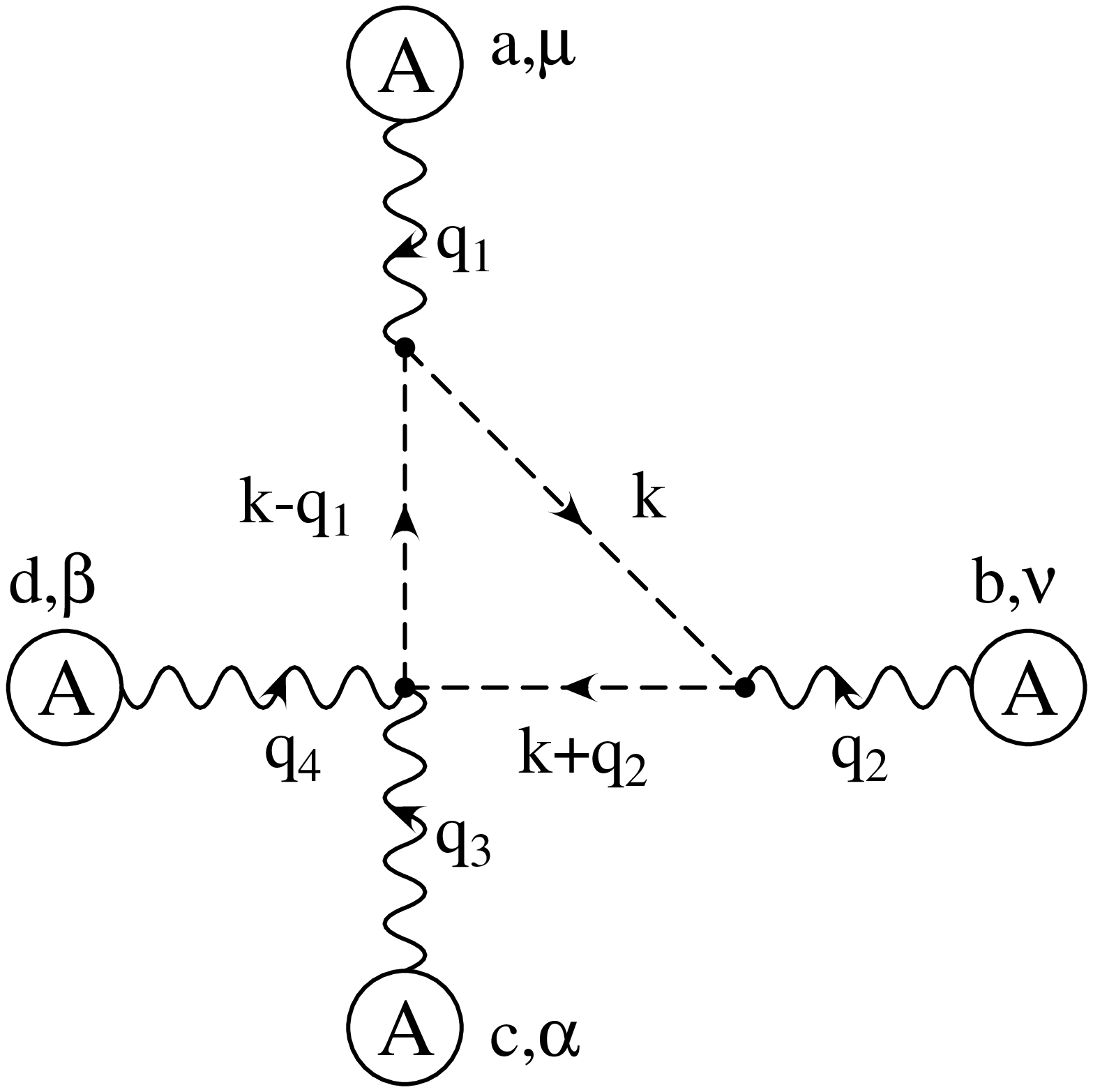}
&
\raisebox{3.6cm}{+ 5 permutations} \\[1cm]
\raisebox{3.6cm}{+} &
\epsfxsize=7cm
\epsffile{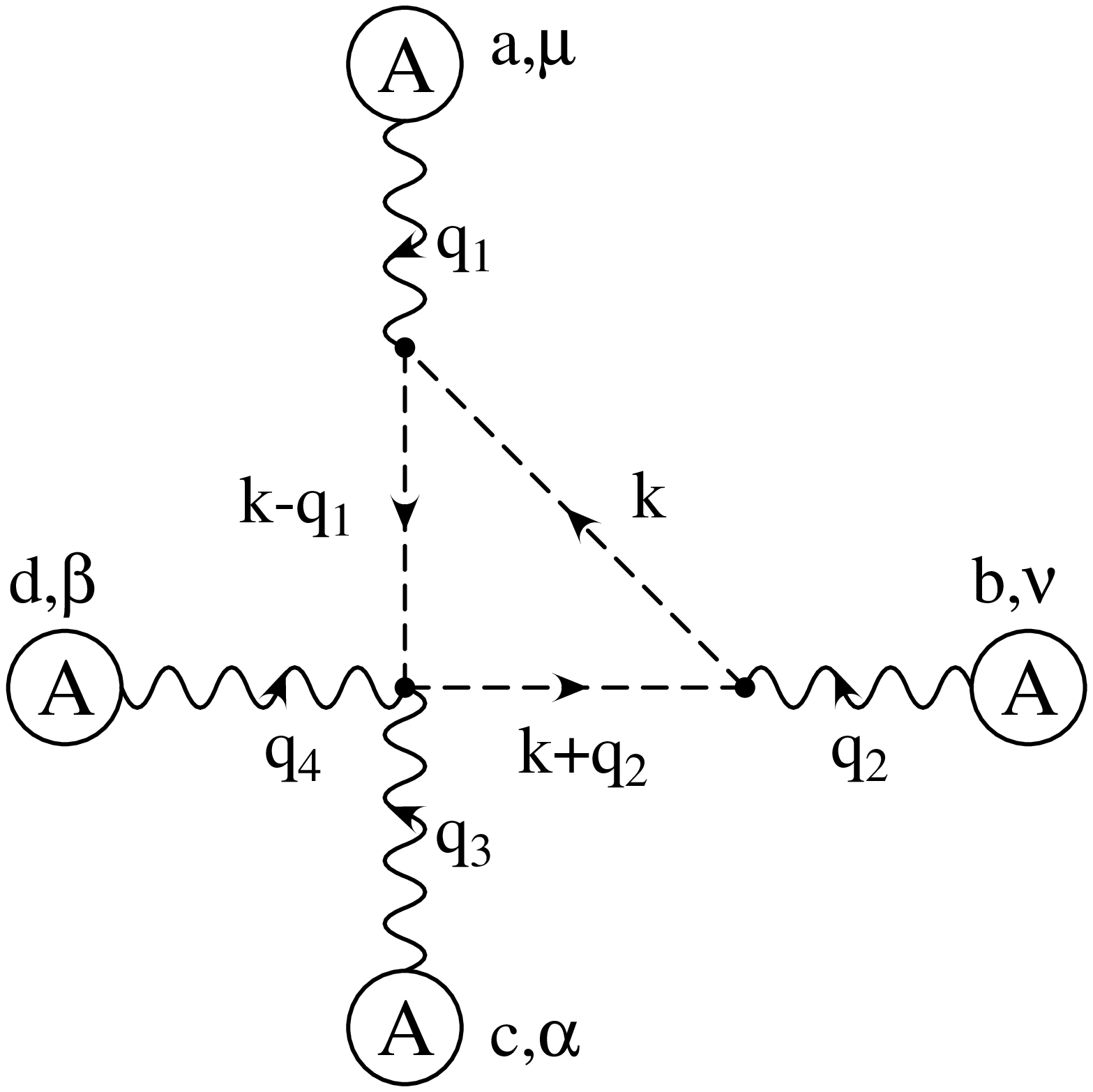}
&
\raisebox{3.6cm}{+ 5 permutations} \\[0.2cm]
\multicolumn{3}{c}{(d)} \\[0.5cm]
\end{tabular}
\end{center}

\newpage
\vspace*{-1cm}
\begin{center}
\begin{tabular}{ll}
\hspace{0.2cm}
\epsfxsize=7cm
\epsffile{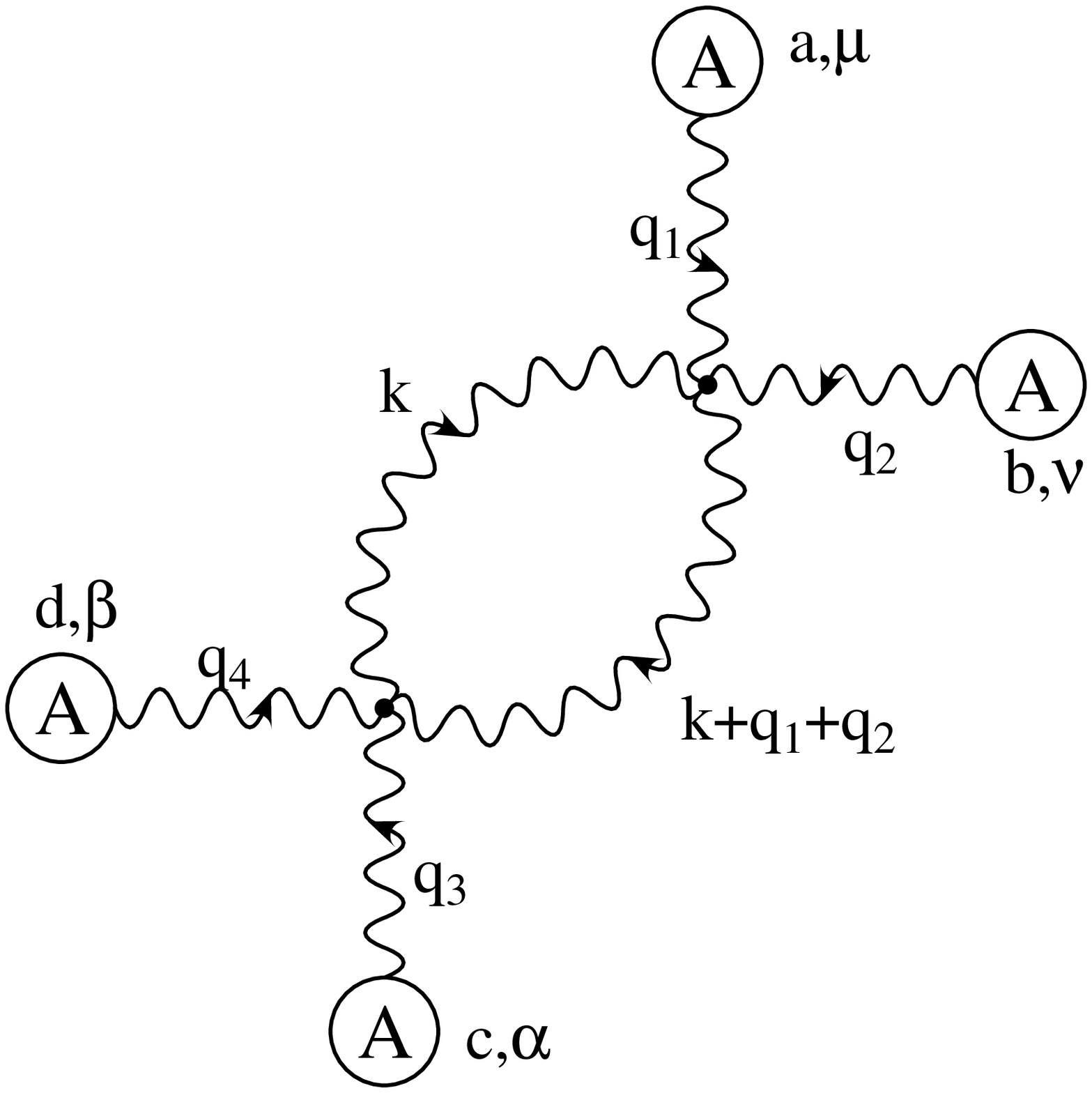}
&
\raisebox{3.2cm}{+ 2 permutations} \\[0.2cm]
\multicolumn{2}{c}{(e)} \\[1cm]

\hspace{0.2cm}
\epsfxsize=6cm
\epsffile{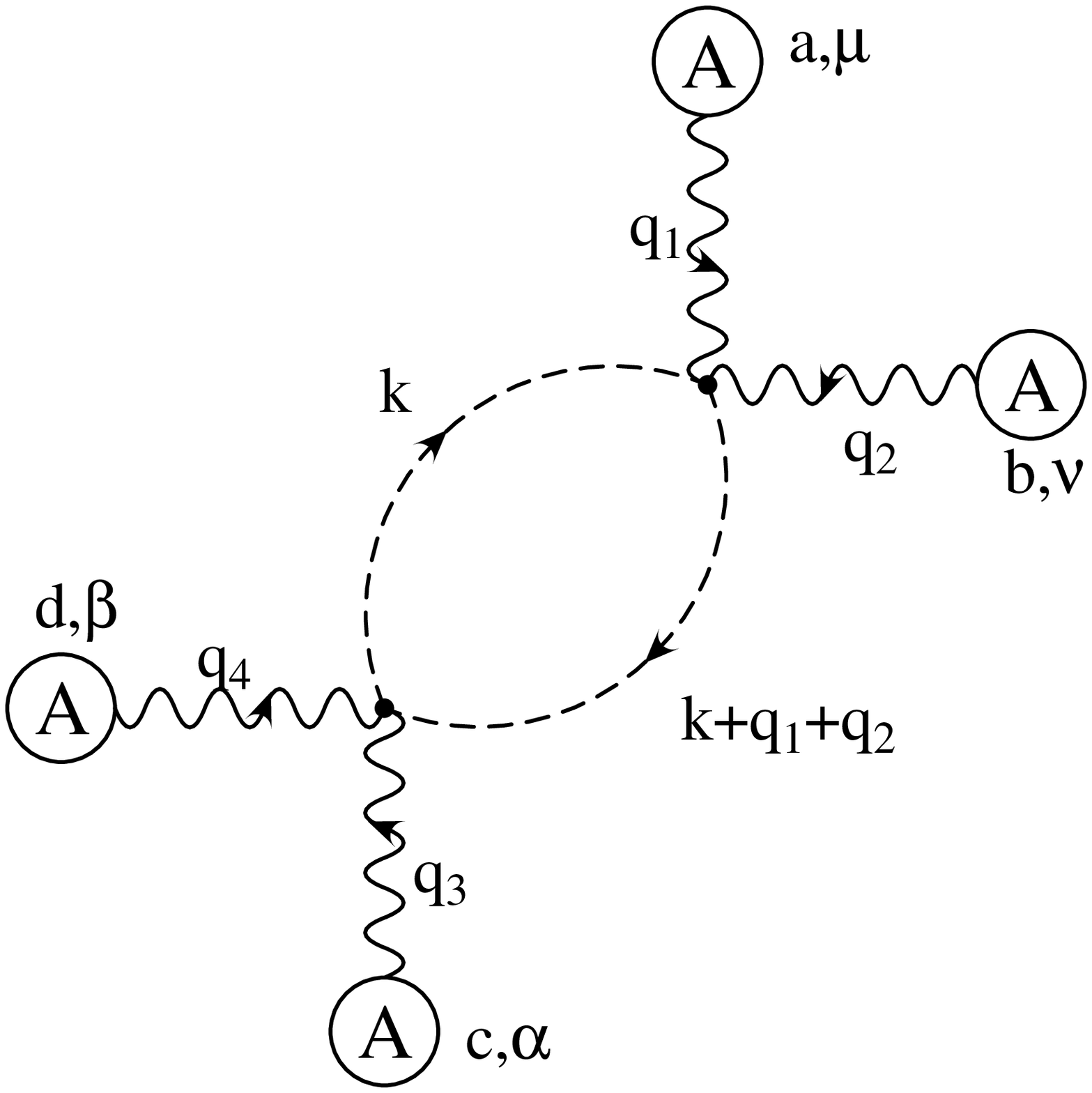}
&
\raisebox{3.2cm}{+ 5 permutations} \\[0.2cm]
\multicolumn{2}{c}{(f)} \\[0.5cm]
\end{tabular}

\Large{Fig.8}
\end{center}


\begin{thebibliography}{99}
\bibitem{defGamma}
     The definition of three-gluon vertex $\Gamma_{\alpha\mu\lambda}$, 
     and therefore, of $\Gamma^F_{\alpha\mu\lambda}$ and 
     $\Gamma^P_{\alpha\mu\lambda}$ in 
     this paper is different from the corresponding ones in 
     Ref.~\cite{rCP} by the overall factor $-1$.
\end{thebibliography}
\end{document}